\documentclass[useAMS,usenatbib]{mn2e}
\usepackage{graphicx,color}
\usepackage[colorlinks=true,citecolor=blue,linkcolor=red,urlcolor=blue]{hyperref}
\usepackage{dcolumn}
\usepackage{url}
\usepackage{times}
\newcommand{\Msun}{\ensuremath{M_{\odot}}}
\newcommand{\Zsun}{\ensuremath{Z_{\odot}}}

\newcommand{\lum}{erg\,s$^{-1}$\,}

\newcommand{\tim}{$\times\,$}

\newcommand{\s}{$\sim\, $}
\newcommand{\de}{$\,^{\circ}$}
\usepackage{amsmath}

\newcommand{\pyr}{$yr^{-1}$}
\newcommand{\pms}{$\pm\,$}
\newcommand{\arcs}{$\arcsec$}

\title[Cavities in Abell~3847]{Detection of a pair of prominent X-ray cavities in Abell~3847}
\author[Vagshette et.al.]{Nilkanth D. Vagshette$^{1}$\thanks{E-mail: nilkanth@prl.res.in}, Sachindra Naik$^{1}$\thanks{E-mail: snaik@prl.res.in}, Madhav. K. Patil$^{2}$\thanks{E-mail: patil@associates.iucaa.in} Satish S. Sonkamble$^{2}$\thanks{E-mail: satish04apr@gmail.com}\\ \\
$^{1}$ Physical Research Laboratory, Navrangpura, Ahmedabad - 380 009, India\\
$^{2}$School of Physical Sciences, Swami Ramanand Teerth Marathwada University, Nanded - 431 606, India.\\
}

\begin{document}
\pagerange{\pageref{firstpage}--\pageref{lastpage}} \pubyear{2016}
\maketitle
\label{firstpage}

\begin{abstract}

We present results obtained from a detailed analysis of a deep \textit{Chandra} 
observation of the bright FRII radio galaxy 3C~444 in Abell~3847 cluster. A 
pair of huge X-ray cavities are detected along North and South directions from 
the centre of 3C~444. X-ray and radio images of the cluster reveal peculiar 
positioning of the cavities and radio bubbles. The radio lobes and X-ray
cavities are apparently not spatially coincident and exhibit offsets by 
\s 61 kpc and 77 kpc from each other along the North and South directions, 
respectively. Radial temperature and density profiles reveal the presence 
of a cool core in the cluster. Imaging and spectral studies showed the removal 
of substantial amount of matter from the core of the cluster by the radio jets. 
A detailed analysis of the temperature and density profiles showed the presence 
of a rarely detected elliptical shock in the cluster. Detection of inflating 
cavities at an average distance of \s55 kpc from the centre implies that the 
central engine feeds a remarkable amount of radio power ($\sim$6.3 $\times 
10^{44}$ \lum) into the intra-cluster medium over \s10$^8$ yr, the estimated 
age of cavity. The cooling luminosity of the cluster was estimated to be 
\s8.30 $\times\, 10^{43}$ \lum, which confirms that the AGN power is 
sufficient to quench the cooling. Ratios of mass accretion rate to Eddington and 
Bondi rates were estimated to be \s0.08 and 3.5 $\times\,10^4$, respectively. This 
indicates that the black hole in the core of the cluster accretes matter through chaotic 
cold accretion.

\end{abstract}


\section{Introduction} \label{sec:intro}
Over the last three decades, observations of many clusters of galaxies showed that 
the X-ray emission is sharply peaked at the position of the central bright galaxy. 
A decrease in the temperature and an increase in the gas density towards the centre have been 
observed, suggesting the existence of cooling flows in the core of clusters  
\citep{1994ARA&A..32..277F}. In the densest central region with high 
cooling rate, the hot gas atmosphere continues to loose energy in the form of 
radiation (mostly in X-rays). The inferred short radiative cooling 
time towards (or at) the centre (shorter than the age of the cluster) indicates
that  the gas located at the centre cools faster than the gas in the outer part 
of the cluster and hence gas flows towards the core to maintain hydrostatic 
equilibrium \citep{1994ARA&A..32..277F,2001ApJ...558L..15B}. 
The excess gas in the central region of the cluster subsonically flows towards 
the core and leads to the formation of cool clouds or stars \citep{1994ARA&A..32..277F}. 
As a result, the central massive galaxies continue to grow with time. The rate of cooling 
flow in nearby clusters is estimated to be in the range of 10 - 100 \Msun\,\pyr \citep
{1994ARA&A..32..277F}. However, in a recent study of massive cluster 
SPT-CLJ2344-4243, the rate of cooling flow was found to be extremely high $\sim$3820 
\Msun\,\pyr \citep{2012Natur.488..349M}.

Present day high resolution (spatial and spectral) observing facilities like 
\textit{Chandra} and \textit{XMM-Newton} have provided us with direct evidences of 
heating of the intra-cluster medium (ICM) in the cores of galaxy clusters through 
active galactic nuclei (AGN) feedback. Positive gradients in temperature profiles and 
short cooling time of the ICM in the core region are the indicative features of 
the cool-core clusters and have been confirmed through several studies in the last decade \citep{1994PASJ...46L..55F,2004A&A...413..415K,2012MNRAS.421.1360H, 2000ApJ...534L.135M,
2002MNRAS.331..273S,2006ApJ...652..216R,2006MNRAS.366..417F,2008ApJ...686..859B,
2010ApJ...712..883D,2010MNRAS.404..180D,2011ApJ...735...11O,2009ApJ...705..624D,
2012AdAst2012E...6G}. From recent studies, it is now established that the majority of cooling 
flow clusters host a powerful radio source associated with the central dominant 
galaxy. High resolution imaging and spectroscopic studies of cool core clusters using 
\textit{Chandra} and \textit{XMM-Newton} observatories revealed
that the interaction of radio jets originating from the core of the cluster with 
the ICM results in formation of cavities, shocks, ripples etc. 
Well known examples which confirm such interactions are the 
Perseus cluster, Abell~2052, Hydra~A  
\citep{2000MNRAS.318L..65F,2001AAS...19916110B,2000ApJ...534L.135M}. 
Thus the bipolar jets from the AGN 
pushes the ambient hot X-ray emitting gas outward and creates X-ray deficient regions 
along the jet direction known as ``cavities". The evidences for the  presence of such
cavities were first reported from the {\it ROSAT} observations of Perseus cluster 
\citep{1993MNRAS.264L..25B}. The availability of high quality data from \textit{Chandra} 
observations of clusters and groups made it possible to detect many such cases and study 
their interaction with the ICM in detail \citep{2000ApJ...534L.135M,2000MNRAS.318L..65F,
2002MNRAS.331..273S,2006ApJ...652..216R,2006MNRAS.366..417F,2008ApJ...686..859B,
2010ApJ...712..883D,2010MNRAS.404..180D,2011ApJ...735...11O,2009ApJ...705..624D,
2012AdAst2012E...6G,2013Ap&SS.345..183P,2012MNRAS.421..808P,2015Ap&SS.359...61S,
2016MNRAS.461.1885V}. Radio observations of several such clusters revealed that 
these cavities spatially match with the radio lobes and are filled with non-thermal 
gas consisting of relativistic particles and magnetic field \citep{2007ARA&A..45..117M,2012AdAst2012E...6G}. Recent studies of cavities suggested 
that the powerful AGN outbursts release enormous amount of energy  ($\sim10^{58} - 
10^{63}\, erg$) and heat the ambient gas in the ICM \citep{2005Natur.433...45M,2007ARA&A..45..117M,2016MNRAS.461.1885V}. 
This amount of energy released (called the AGN feedback) is sufficient 
to balance the cooling of hot atmosphere surrounding the central engine 
\citep{2004ApJ...607..800B,2006MNRAS.368L..67B,2006MNRAS.373..959D,
2006ApJ...652..216R,2009AIPC.1201..301B,2010ApJ...712..883D,
2016MNRAS.461.1885V}. Apart from AGN feedback, there are other 
possible heating mechanisms as proposed by several authors; 
like heating due to merging \citep{2006PhR...427....1P,
2007PhR...443....1M}, supernovae heating \citep{2004A&A...425L..21D}, 
cosmic rays heating \citep{2004A&A...413..441C,2004PhRvL..92s1301T}, 
etc. However, AGN feedback mechanism provides the most acceptable  
explanation for heating the ICM.

3C~444 has been identified as a brightest galaxy at the centre of Abell~3847 
cluster \citep{1965AJ.....70..384W} and belongs to the class of weak-line FRII 
radio galaxy \citep{2002MNRAS.330..977T,2012ApJ...745..172D} positioned at 
RA=22:14:25.7, DEC=-17:01:36 at a redshift of $z$=0.153. It is one of the 
famous radio sources and has been widely studied in radio bands 
\citep{1979crs..book.....K,1981AJ.....86.1306G,1986A&AS...65..485R,
2008ApJ...678..712D,2009ApJ...694..268D}. However, the X-ray properties 
of the cluster have not been investigated due to the lack of observations. 
Using 20 ks \textit{Chandra} observation, \cite{2011ApJ...734L..28C} identified 
the large scale shock features in the cluster. Apart from these results, nothing
has been reported on the morphological and spectral properties of hot gas in the
cluster. In the present work we report the imaging and spectral characteristics of hot gas 
and compare it with the radio morphology.

In this paper, we present results obtained from a systematic study of 3C~444 in 
Abell~3847 by using 166 ks {\it Chandra} archival observations. The structure
of the paper is as follows: Section 2 provides details on the \textit{Chandra} observation 
of the cluster and data reduction strategies. Section 3 describes results obtained 
from X-ray imaging analysis and its comparison with the radio map. We also provide 
spatially resolved spectral analysis of hot gas and the detection of shock in this 
section. Section 4 reports a brief discussion on cavity energetics, heating and 
cooling mechanism and central accretion phenomena. Finally, Section 5 presents the
conclusions drawn from the study. Throughout this paper, we use cosmology with 
$H_0 = 70\, km s^{-1} Mpc^{-1}$, $\Omega_M$=0.3 and $\Omega_{vac}$=0.7 for 
3C~444 cluster. This corresponds to an angular scale of 
2.66 kpc arcsec$^{-1}$.


\section{Observation and Data reduction}

3C~444 was observed with {\it Chandra} ACIS-23678 detector on 12-13 October 
2013 with a net integration time of $\sim$166 ks (ObsID~15091) with the source 
focused on the back illuminated ACIS-S3 chip in `VFAINT' mode. In this study, we 
used archival data of 3C~444 obtained from \textit{Chandra} X-ray Center
(CXC)\footnote{\color{blue}{{http://cda.harvard.edu/chaser/}}}. The \textit{Chandra} 
data were reprocessed following the standard data reprocessing procedure as 
described in \textit{Chandra} interactive analysis of observations (CIAO-4.6\footnote
{\color{blue}{{{http://cxc.harvard.edu/ciao/threads/index.html}}}}) and employing 
CALDB V 4.6.2. To identify the flaring events during 3C~444 observation, we initially 
removed bright point sources from the raw image and than extracted X-ray counts in 
2.5-7 keV energy range \citep{2009ApJ...705..624D}. The \textit{lc$\_$sigma$\_$clip} 
script of \textsc{CIAO-sherpa} package was used to identify and mark detected flares 
with 3$\sigma$ clipping threshold. However, during the \textit{Chandra} observation
of 3C~444, no significant flaring events were evident in the light curve.

ACIS-S blank-sky background data was used for background subtraction from the 
science data. CIAO task \textit{$acis\_bkgrnd\_lookup$} was used to identify the 
blank-sky background file corresponding to 3C~444 observation. Before this, the 
background data was normalized in such a way that the count rate of blank-sky and 
science data in 9--12 keV range matched, which was done by using the method described 
by \cite{2001ApJ...551..160V} and \cite{2009ApJ...705..624D}. Identification of  
point sources in the detector field of view was done by using the CIAO script 
{\it wavdetect} and were excluded from further analysis. The spectra and 
response files were generated by using {\it specextract} task of CIAO and 
XSPEC-12.8.1 package of HEASOFT was used to perform spectral fitting.

\section{Results}
\subsection{Imaging of radio galaxy 3C~444 in A3847}

\begin{figure*}
\vbox{
\includegraphics[trim=00 .7cm 00 00, clip=yes, width=8cm,height=7cm]{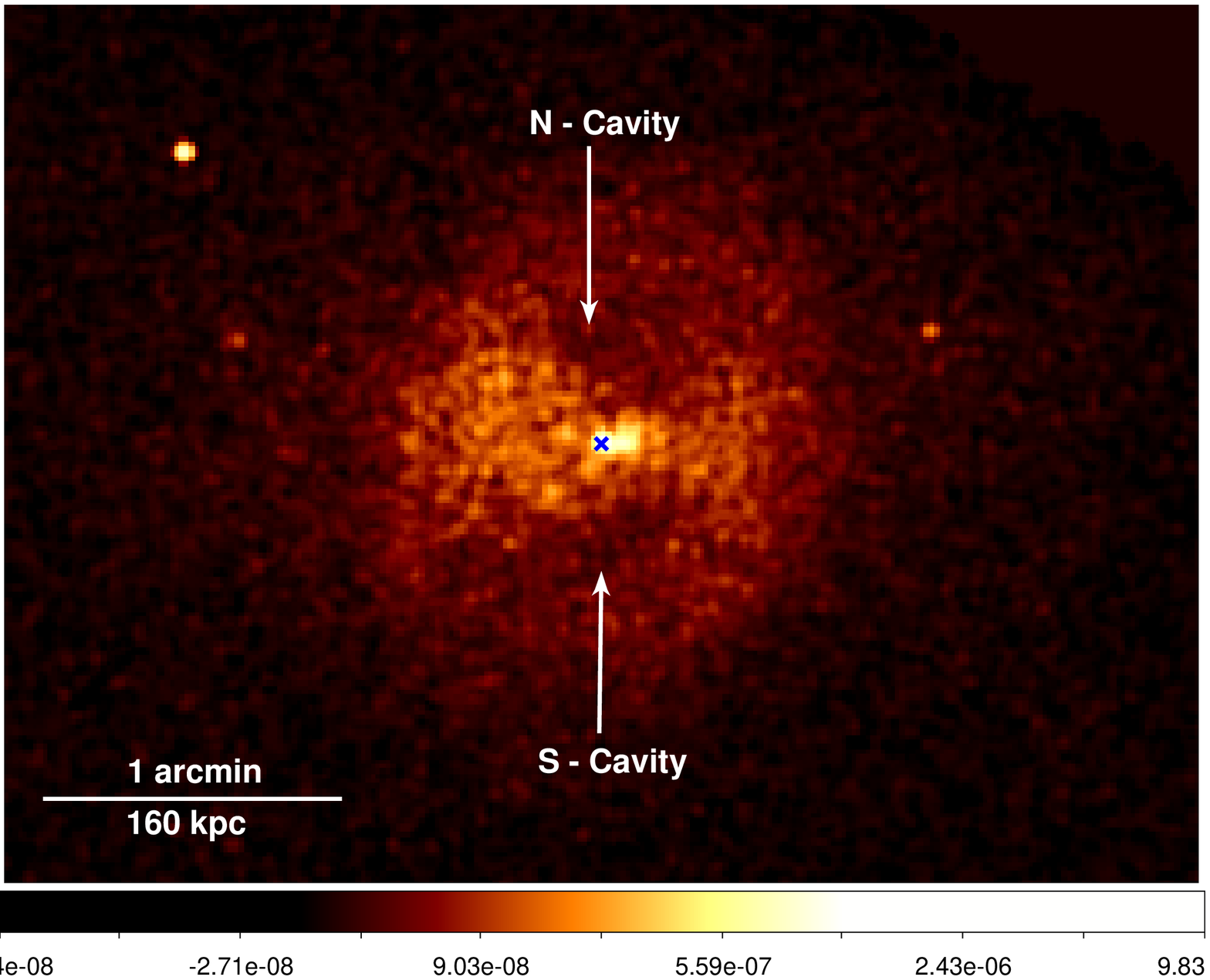} 
\includegraphics[width=8cm,height=7cm]{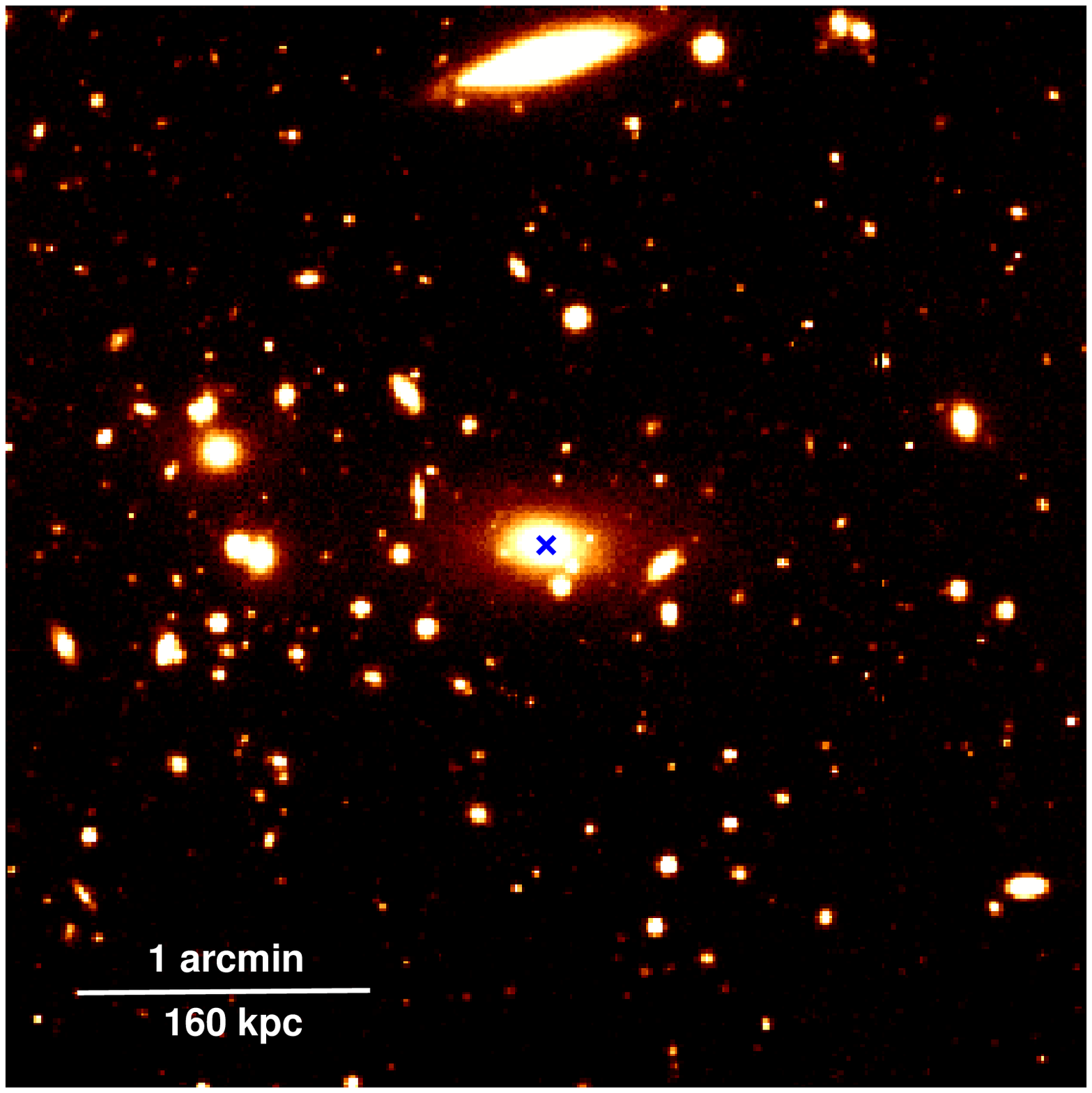} } 
\caption{Background subtracted and  exposure corrected $4'\times 4'$ \textit{Chandra} X-ray and 
R-band {\it Gemini-South} $4' \times 4'$ images of A3847 cluster are shown in the left and
right panels, respectively. The arrow marks in the {\it Chandra} X-ray image indicate the 
presence of X-ray deficient regions in the cluster. In the image, North and East are in 
upward and left directions, respectively.}
\label{fig1} 
\end{figure*} 

A 2 pixel (1$\arcsec$) Gaussian smoothed, background subtracted and exposure-corrected 0.3-6 keV 
band $4' \times 4'$ {\it Chandra} X-ray image of A3847 is shown in the left panel 
of Figure~\ref{fig1} whereas the right panel depicts the R-band optical image of
the cluster obtained from the archive of {\it Gemini-South} observatory. The 
``X''-marks in both the images denote the approximate centre of the bright  
galaxy 3C~444. The images shown in Figure~\ref{fig1} suggest the presence of 
diffuse X-ray emission with several peculiar features in the cluster though 
the central bright extended source is only seen in the optical image. The 
orientation of the central bright galaxy in the optical image is along the 
East-West direction whereas the X-ray morphology is found to 
be peculiar due to the presence of X-ray deficient regions. In the \textit{Chandra} 
X-ray image, arrow marks along the North-South direction show the X-ray deficient 
regions in the cluster and have been visualized clearly using different techniques as
discussed below.

\begin{figure*}
\includegraphics[width=8cm,height=7cm]{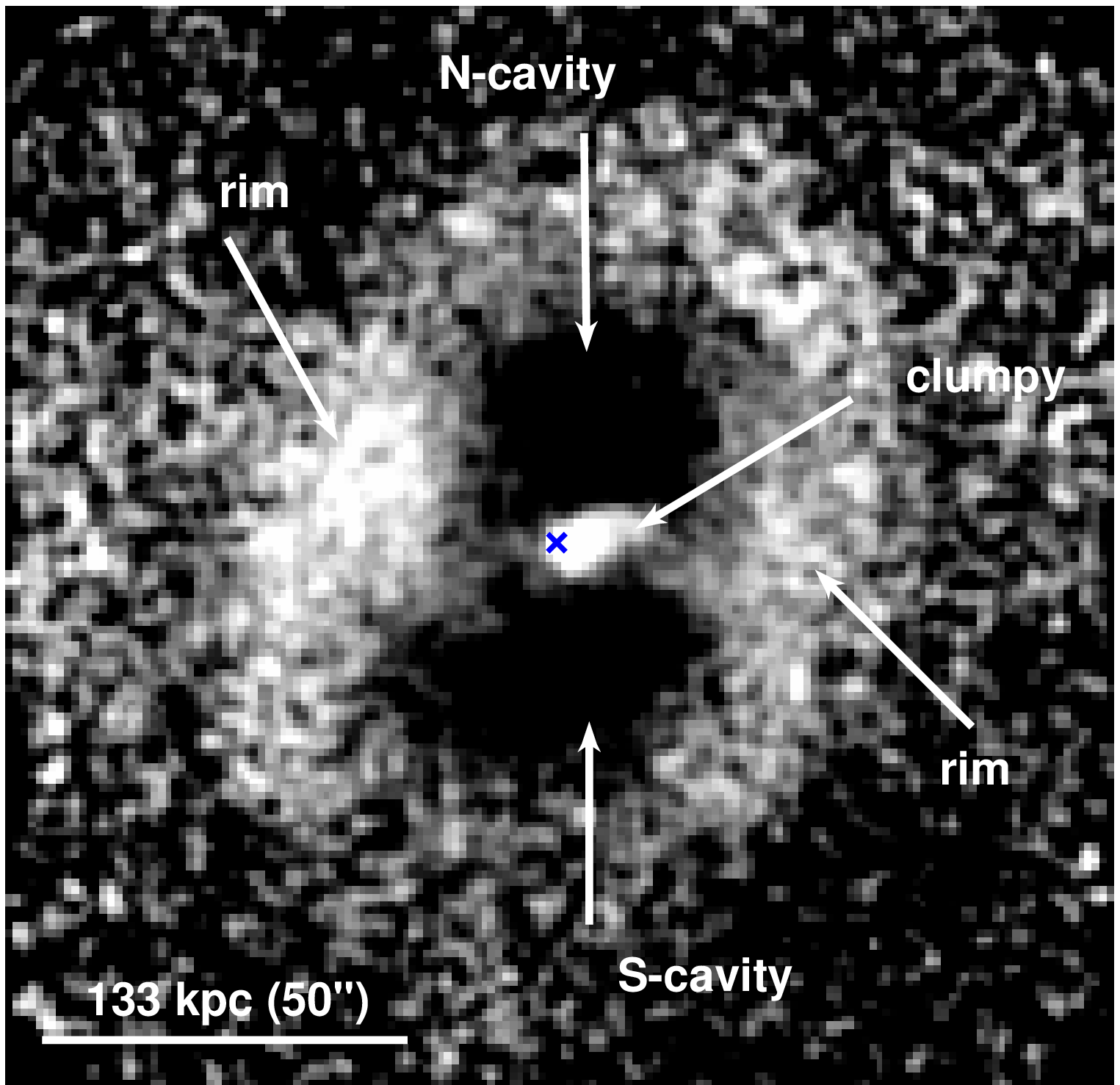} 
\includegraphics[width=8cm,height=7.01cm]{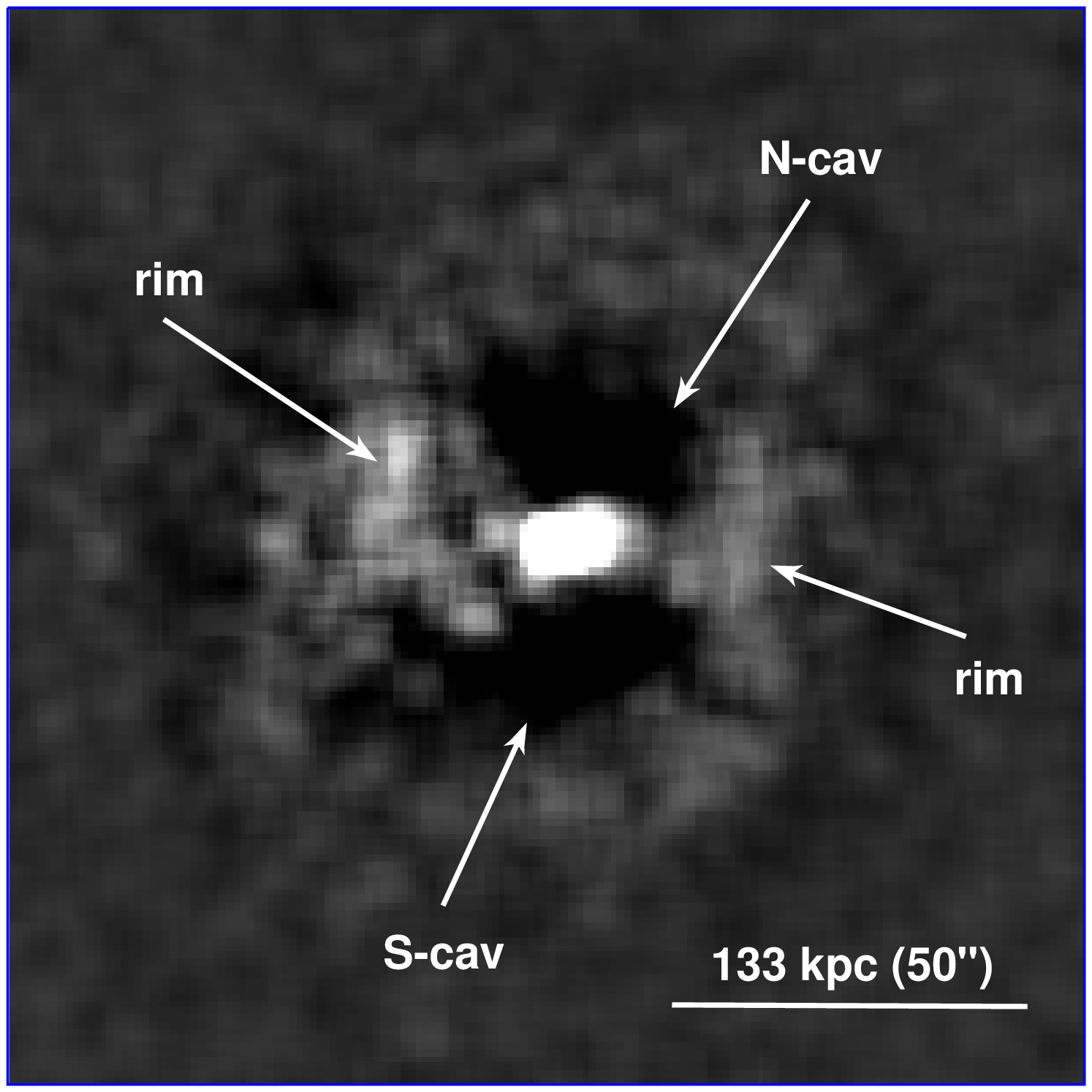}
\caption{0.3-6 keV quotient image for central 2.5$\arcmin$ x 2.5$\arcmin$ field
derived by dividing the original image with the 2D-$\beta$ model image (left panel), whereas
the right panel shows the residual image generated by unsharped mask
imaging technique.}
\label{fig2} 
\end{figure*}

To identify the position and size of the X-ray deficient regions accurately and 
enhance the hidden features in the central part of A3847, we generated quotient and residual 
maps of the cluster by using 2D-$\beta$ model \citep{2010ApJ...712..883D} and 
unsharp mask imaging \citep{2010ApJ...712..883D,2009ApJ...705..624D} techniques. 
In 2D $\beta$-model, the smoothed image of the cluster was produced by fitting 
2D $\beta$-model on the 0.3-6 keV \textit{Chandra} image. The resultant best 
fit parameters of these functions are slope $\alpha$ = 0.98$\pm$0.02 and core 
radius r$_0$ = 54.41$\pm$1.4 arcsec. The original image was then divided with
the smoothed 2D $\beta$-model image. The resultant quotient image is shown in the left 
panel of Figure~\ref{fig2}. Using the unsharp mask technique, the residual image of the 
cluster was obtained by subtracting 20 pixel (10$\arcsec$) Gaussian smoothed 0.3-6 keV image 
from 3 pixel (1.5$\arcsec$) Gaussian smoothed image in the same energy band and is shown in the 
right panel of Figure~\ref{fig2}. Both the quotient and residual images revealed the 
presence of a pair of giant X-ray deficient regions, known as cavities, along the 
North-South direction of the X-ray peak. These cavities are found to be surrounded 
by an excess X-ray emission (bright rims). Apart from this, the central nuclear 
clumpy region is found to be elongated towards East from the X-ray centre (indicated 
with an ``X'' mark).

\begin{figure*}
\centering
\includegraphics[width=8cm,height=7cm]{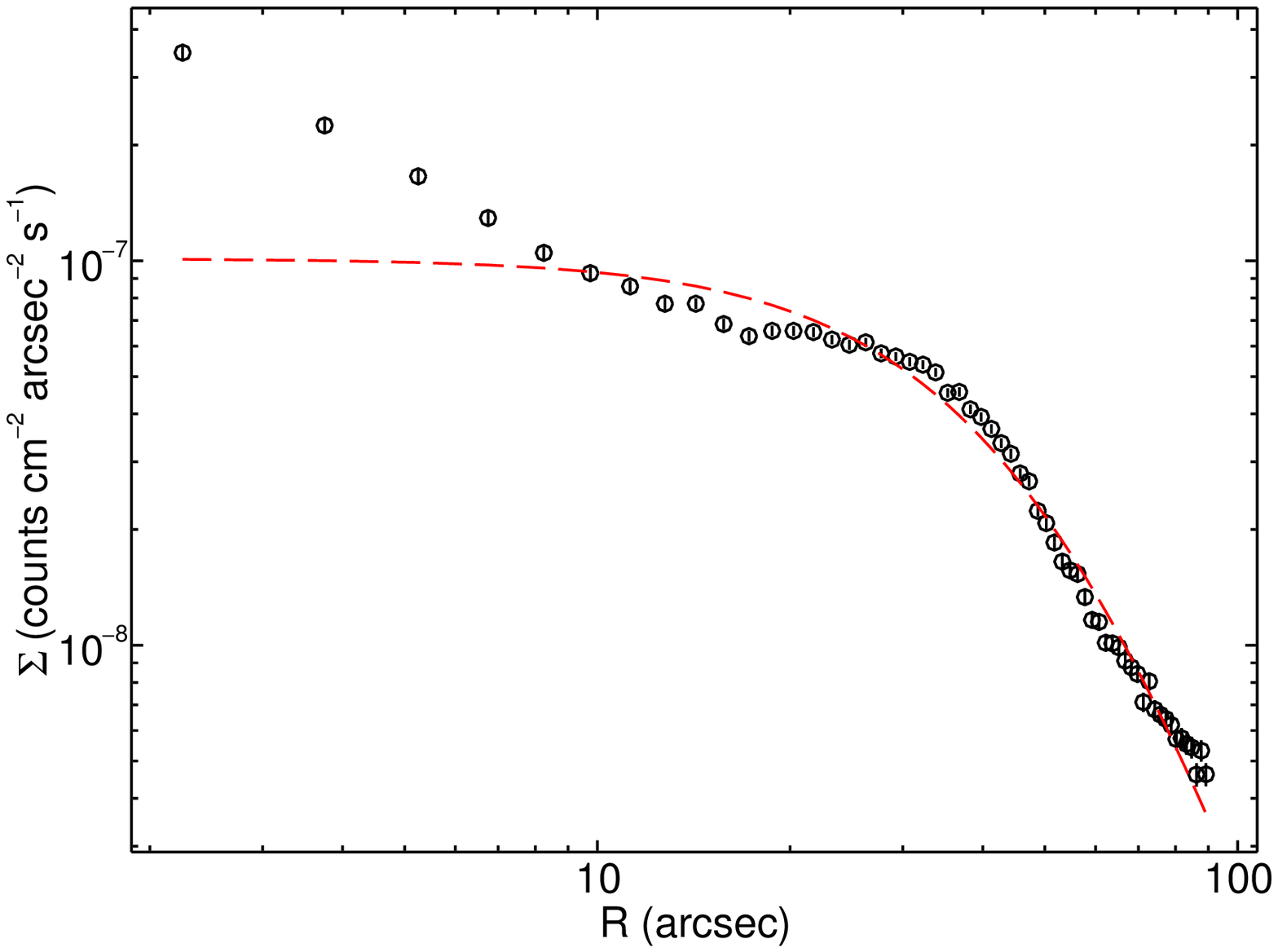} 
\includegraphics[width=8cm,height=7cm]{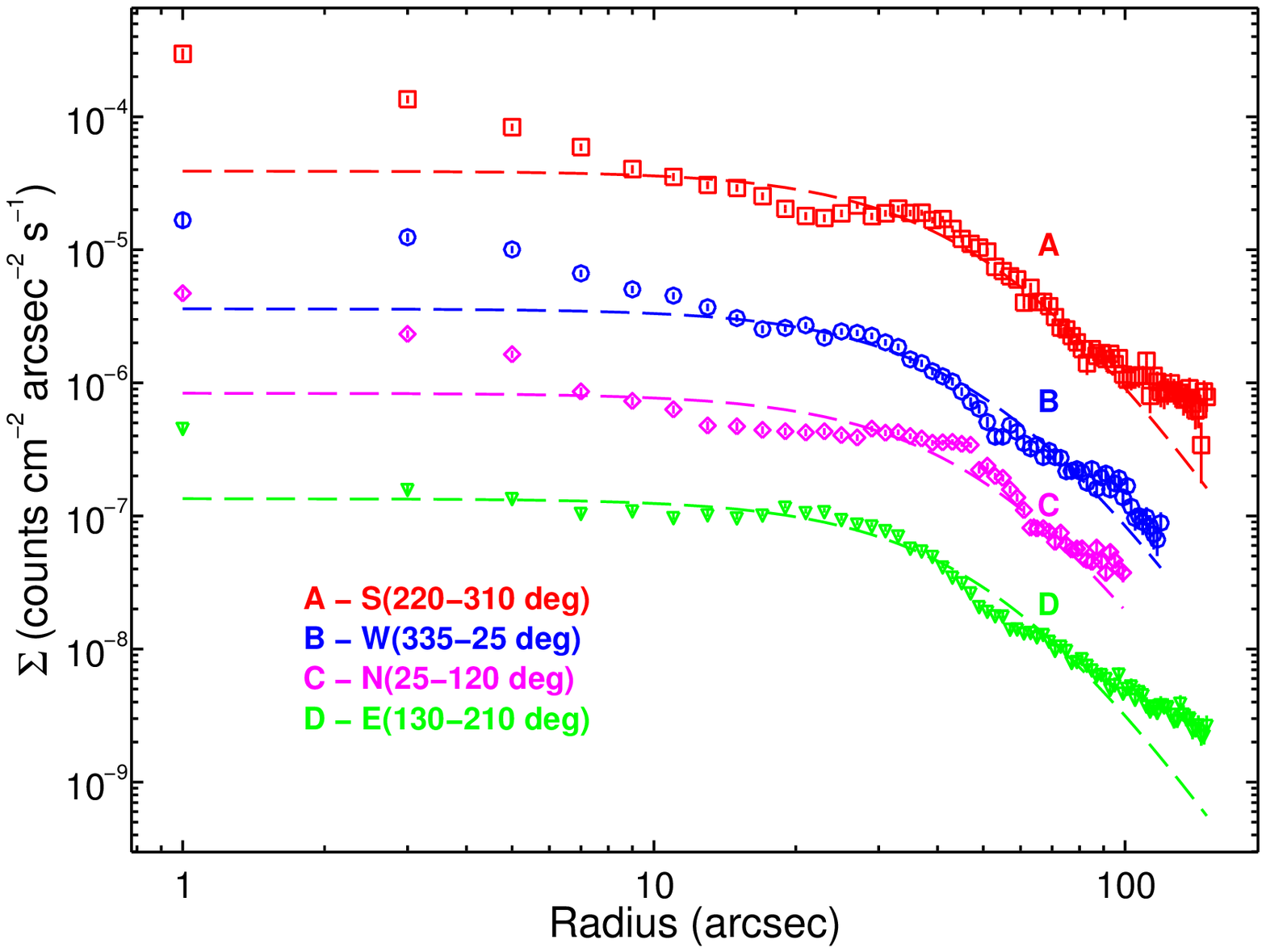} 
\caption{Surface brightness profiles: \textit{Left panel} shows the 
azimuthally averaged radial surface brightness profile fitted with 
1D $\beta$-model whereas the \textit{right panel} shows the radial 
surface brightness profiles extracted from four different wedge shaped 
sectors.}
\label{fig3} 
\end{figure*} 
 
To investigate the distribution of X-ray emitting hot plasma in A3847 cluster, 
surface brightness profiles were extracted in 0.3-6 keV range by considering 
annular regions centered on the central peak position (for more details see 
\citealt{2016MNRAS.461.1885V}). The resultant surface brightness plots, corrected 
for background and exposure, are shown in Figure~\ref{fig3}. Assuming ICM 
of A3847 to be in hydrostatic equilibrium in the gravitational potential
of central dominant galaxy and well described by King's approximation to the 
isothermal sphere \citep{2012AdAst2012E...6G,1962AJ.....67..471K}, the surface 
brightness profiles were fitted with the 1D $\beta$-model through the 
expression
 \begin{equation}
\Sigma_{(r)} = \Sigma_{(0)} \left[1 + \left(\frac{r}{r_0}\right)^2 \right]^{-3\beta+0.5}
\end{equation}

where, the functional parameters $\Sigma_{(0)}$, $r_0$ and $\beta$ represent
the central brightness, core radius, and slope, respectively. Best-fit values 
obtained from 1D $\beta$-model fitting are $\beta$ = 1.04$\pm$0.04 and 
$r_0$ = 56.21$\pm$1.75 arcsec. A plot of azimuthally averaged surface 
brightness profile (black circle), fitted with 1D $\beta$-model (dash line), 
is shown in the left panel of Figure~\ref{fig3}. The azimuthally averaged 
surface brightness profile exhibits varieties of features such as (i) excess 
emission from the central region compared to the $\beta$ model confirming that 
A3847 to be a cool core cluster, (ii) X-ray depression between 10--25 
arcsec region suggesting the presence of cavities, (iii) the excess 
emission at 25--45 arcsec and (iv) rapid fall of surface brightness at 
larger radii (above 50 arcsec).

To investigate the surface brightness features along the cavity (North and South 
directions) and non-cavity (East and West directions) regions, we derived 
surface brightness profiles in 0.3--6 keV band from four different wedge shaped 
sectors, covering North (25\de--120\de), East (130\de--210\de) South (220\de--310\de) 
and West (335\de--25\de). The background subtracted and exposure corrected 
resultant sectoral surface brightness profiles are shown in the right panel 
of Figure~\ref{fig3}. For clarity and better presentation, surface brightness
values obtained for four sectors are arbitrarily scaled along the vertical axis. 
For comparison, the best-fitted azimuthally averaged 1D $\beta$-model is 
over-plotted in dashed lines (with appropriate normalization). These profiles 
clearly show the deficiency in surface brightness values compared to the 
$\beta$ model in annular regions between 10--30 arcsec radii along the North 
and South directions. All the profiles show excess emission at the centre 
of the cluster. However, between 70--80 arcsec annular regions, the surface 
brightness profiles along the North and South directions show shock like 
features and are discussed in detail in Section 3.6.

\subsection{Spectral analysis}

The source and background spectra were extracted by following the procedure 
described in \cite{2016MNRAS.461.1885V}. For background estimation, 
blank-sky background observation was used. As A3847 is an X-ray bright cluster 
and the \textit{Chandra} observation used in the present work is of very long
exposure, spectral analysis of data extracted from the several regions of interest
is viable. Spectra for such regions like nucleus of the cluster (within central
2\arcs radius), cavities (along North and South directions), shock regions
etc. were extracted for model fitting. As the X-ray emission from galaxy clusters 
is dominated by thermal Bremsstrahlung, spectral fitting were performed by using 
collisionally ionized plasma code APEC \citep{2001ApJ...556L..91S} and absorption 
(wabs) fixed at the Galactic value. All the extracted spectra were fitted in
the 0.5 -- 7 keV energy band.

\subsection{Cavity environment}

\begin{figure*}
\centering
\vbox{
\includegraphics[width=8cm,height=7cm]{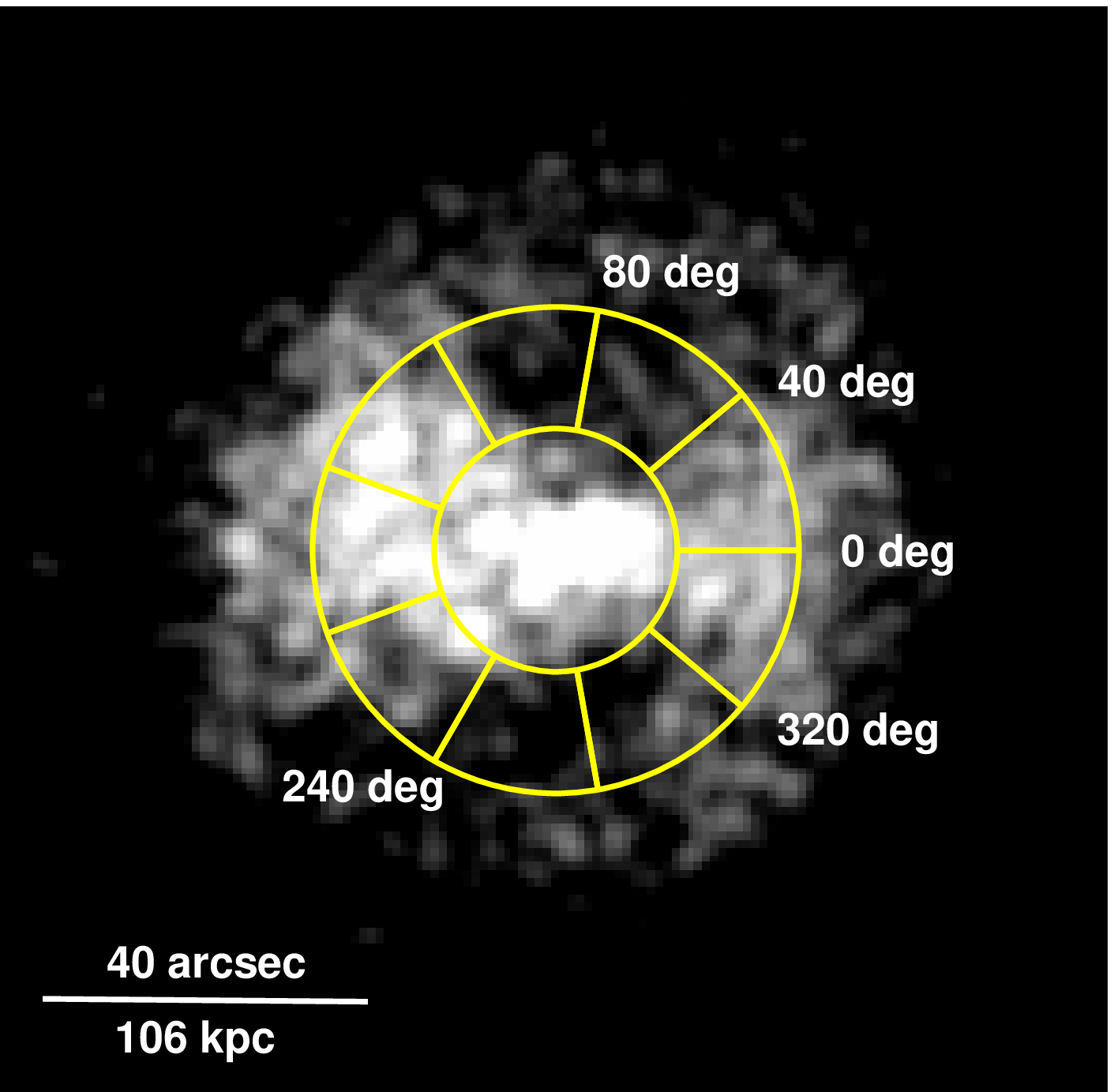}
}
\includegraphics[width=8cm,height=7cm]{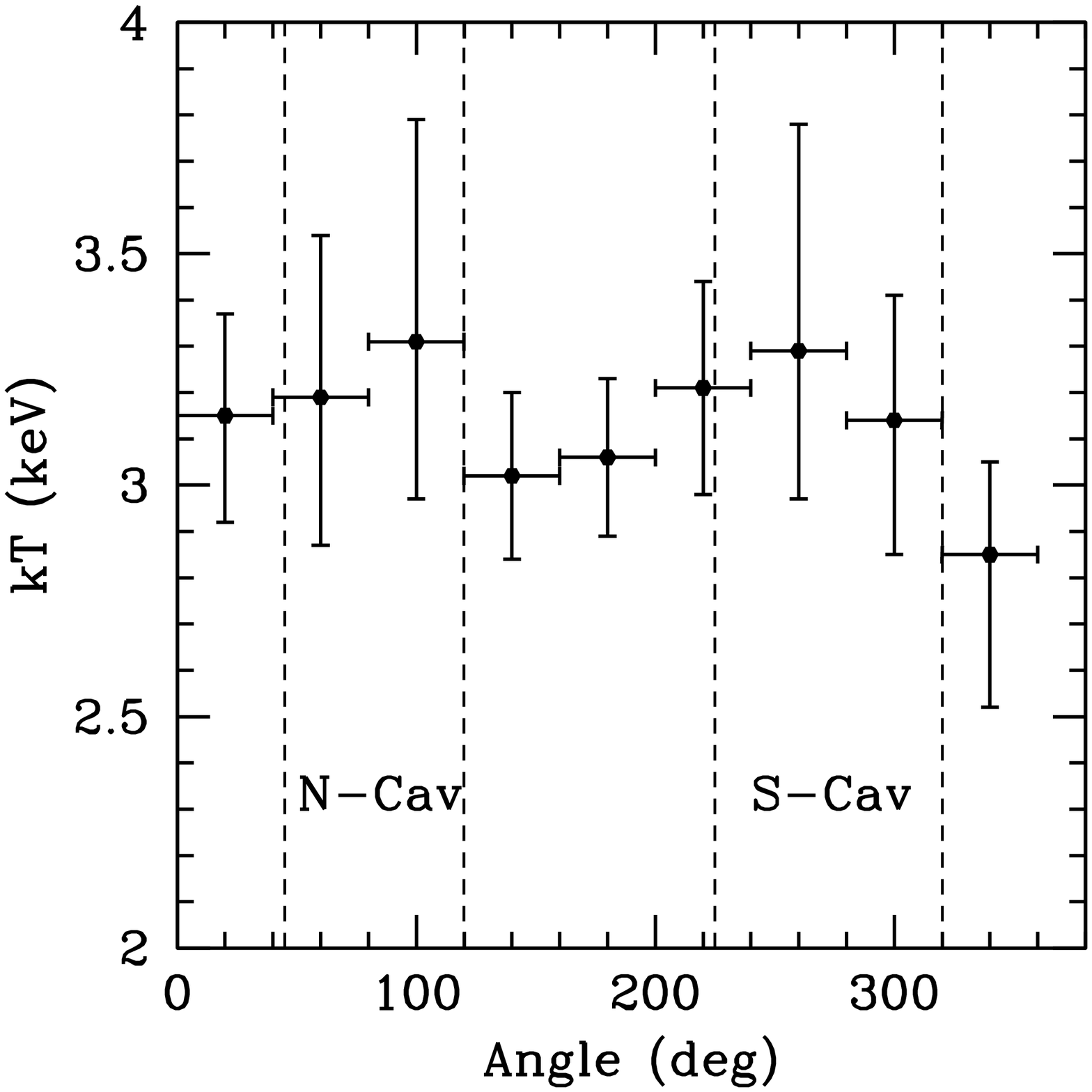}
\includegraphics[width=8cm,height=7cm]{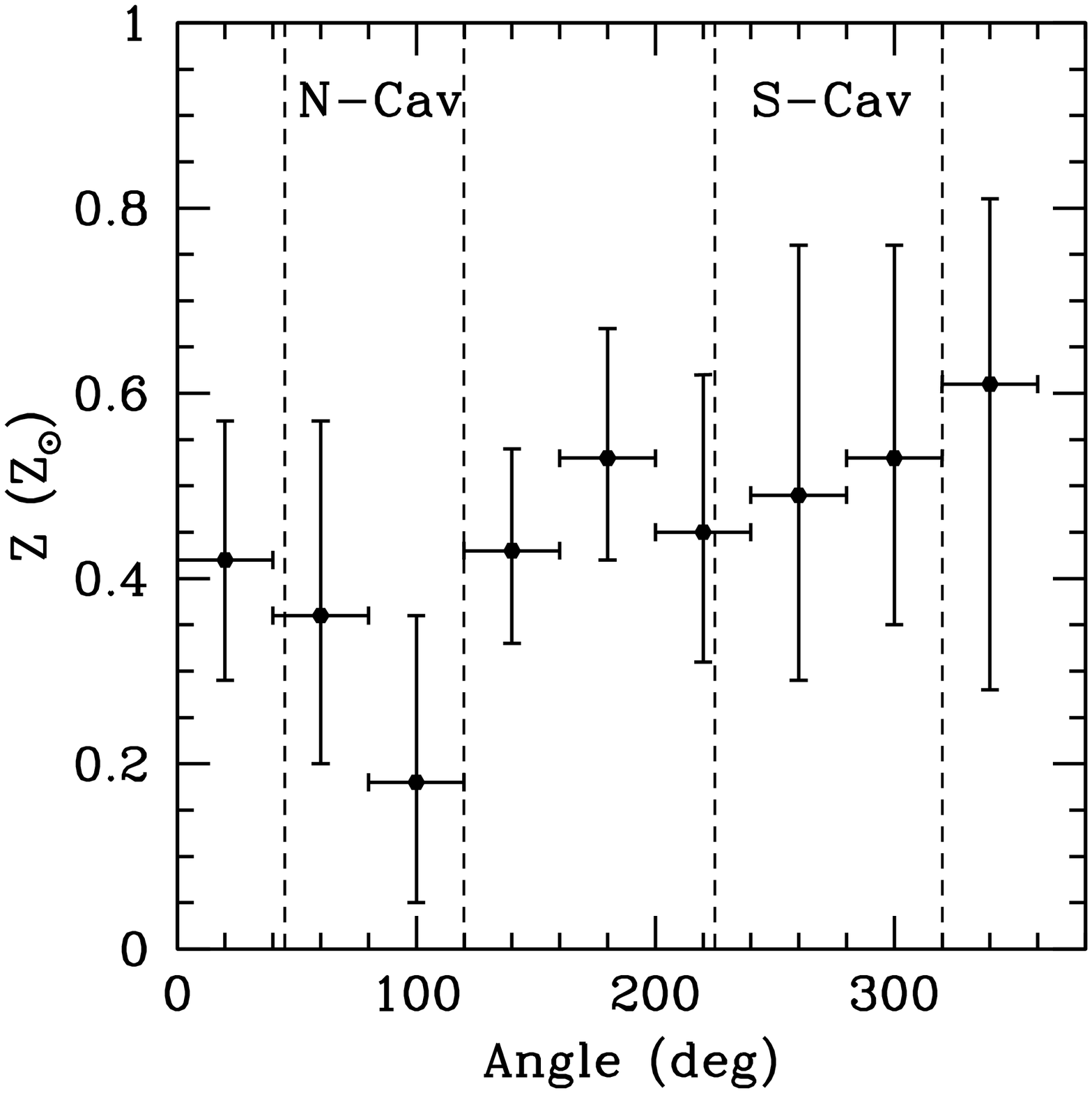}
\caption{\textit{Top panel} shows the sectors in the annular region of 15--35 arcsec radii 
from which spectral extraction was carried out separately. \textit{Bottom panel} shows 
the variations in temperature (left panel) and metal abundance (right panel) in the X-ray deficient 
(cavity) and surrounding regions, respectively.}
\label{fig4}
\end{figure*}

The presence of cavities in A3847 are clearly evident in the residual images of the
cluster (Figure~\ref{fig2}) and  have also been confirmed in the surface brightness 
profiles (Figure~\ref{fig3}). We made an attempt to perform a comparative study of the 
cavities and its environment by fitting spectra obtained from different sectors 
of the ICM enclosed within an annular region of 15--35 arcsec radii which 
was divided into nine equal sectors (Figure~\ref{fig4}, top panel). Spectral 
extraction was carried out from each sector separately. The spectral parameters 
like temperature and abundance were obtained by fitting each spectrum with a 
single temperature model. Figure~\ref{fig4} shows the variations in temperature 
(bottom left panel) and metal abundance (bottom right panel) values which reveal 
that the temperature of X-ray deficient regions (cavities) in A3847 was relatively 
higher than its surrounding regions, while the metal abundance was lower or comparable 
with the surrounding ICM.

\subsection{Thermodynamical properties of the hot gas in ICM}

\begin{figure}
\centering
\includegraphics[width=8.cm,height=14.cm]{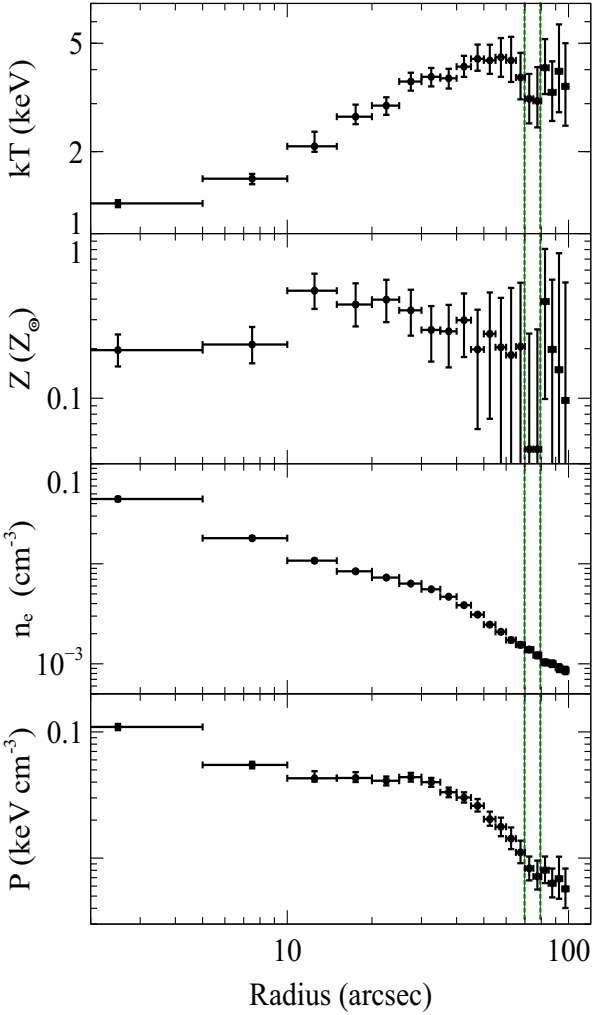} 
\caption{Thermodynamical parameters estimated from spectral fitting of
data obtained from azimuthally averaged concentric annular regions from 
cluster centre to 100 arcsec radii with bin-width of 5 arcsec. Top to 
bottom panels show the temperature, abundance, electron density 
and pressure profiles, respectively.}
\label{fig5} 
\end{figure}

To investigate the thermodynamical properties of hot gas in the ICM of 
A3847, we extracted azimuthally averaged projected spectra from circular 
annuli centred on the X-ray peak up to 2 arcmin radii with each bin width 
of 5 arcsec. These spectra were extracted by assuming the ICM of the cluster 
as spherically symmetric. Blank-sky observation, as mentioned earlier, was 
used for background subtraction. Spectral fitting was performed by using 
single temperature collisionally ionized plasma code (APEC) with fixed 
Galactic absorption. All the three parameters (temperature, abundance and 
normalization) were allowed to vary during the fit. Once the values
of temperature and normalization were obtained, other thermodynamical 
parameters such as electron density ($n_e$) and pressure ($p$) were 
estimated by following the method described in \cite{2012AdAst2012E...6G,
2016MNRAS.461.1885V}. Azimuthally averaged projected temperature, 
metallicity, electron density and pressure profiles are shown in the
top to bottom panels of Figure~\ref{fig5}, respectively. The rising 
temperature and decreasing metallicity, density and pressure in the 
outward direction from the centre indicate that A3847 is a cool core 
cluster. In the temperature profile, the minimum and maximum values 
were found to be $\sim$1.2 keV (at the centre) and $\sim$ 5 keV (in 
50-60 arcsec annular region), respectively. However, in the 70-80 arcsec 
annular region, a sharp drop in the temperature by a factor of $\sim$2 is clearly seen. 
Similar effects are also seen at the same position in the metallicity and pressure 
profiles. Though the change in the number density was not significant
like other parameters, a marginal decrease in the number density 
is evident in the third panel of Figure~\ref{fig5}.
  
\subsection{Sectoral spectral analysis}

\begin{figure*}
\centering
\includegraphics[width=16.cm,height=14.cm]{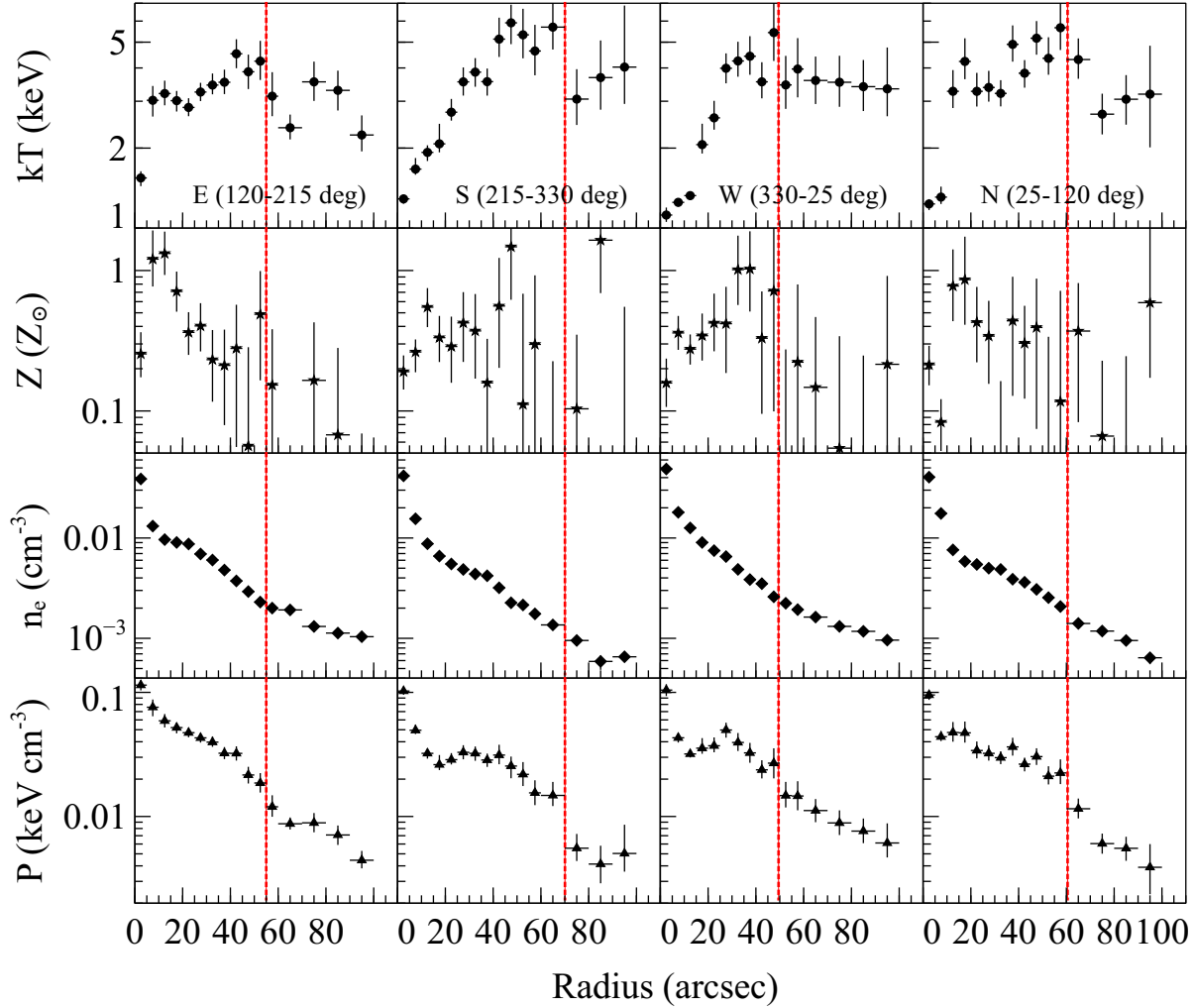} 
\caption{Thermodynamical parameters derived from the analysis of spectra 
extracted from four sectors (East, South, West and North). Top to bottom 
panels show the temperature, abundance, electron density and pressure profiles, 
respectively. The vertical dashed lines represent the expected shocked regions 
at corresponding radii.}
\label{fig6} 
\end{figure*}

To study the nature of gas distribution in different directions, we extracted 
spectra from four different sectors (East, South, West and North) by selecting 
annular regions, each of width of 5\arcs\, (up to 60\arcs) and 10\arcs 
(from 60\arcs to 100\arcs) radius from the centre of the cluster. All the spectra 
for each of the four sectors were well fitted with a single temperature (APEC) 
model and Galactic absorption yielding the values of metallicity and temperature at
different radial distances. Using the parameters obtained from spectral fitting, 
we estimated electron density and pressure at different radii for each of the sector. 
The resultant temperature, metallicity, electron density and pressure profiles 
for all the four sectors are plotted in the top to bottom panels of Figure~\ref{fig6}, 
respectively. The temperature profiles revealed that the ICM temperature at
certain radii from the centre of A3847 is different in different sectors. 
Along North and South directions, at a radial distance of $\sim$70 arcsec 
(\s180 kpc) from cluster centre, the temperature profiles showed a sharp drop 
by a factor of $\sim$2. However, along East and West directions, the temperature
drop by a factor $\le$2 was observed at $\sim$50 arcsec (\s133 kps) 
from the centre (top panels). A similar drop was  also   evident in the pressure
profiles (bottom panels) though the abundance and electron density profiles showed 
marginal drops in the corresponding values (second and third panels). These observed 
changes suggest the presence of shocks at the radial distance of $\sim$70 arcsec
in North and South directions and $\sim$50 arcsec in East and West directions
as marked with dashed lines in the figure.

Though the error bars in the metallicity plot in Figure~\ref{fig6} (second panels) 
are large, a careful observation of the profiles for the four sectors provides an 
important information. Metal abundance profiles show that the distribution of 
metallicity in the cluster is not symmetric in all the four directions. In the East 
sector, metallicity profile peaks sharply (1.3 Z$_\odot$) at a radial distance of 
\s15 arcsec radius (40 kpc) beyond which it decreases monotonically, while in the 
West sector it gradually increases to a peak value of 1~Z$_\odot$ at a distance of 
30--40 arcsec radii (80--107 kpc) beyond which it again decreases. Gradual increase 
in the value of metallicity from the centre to \s15 arcsec in the East sector and \s30 
arcsec radius in the West sector can be explained as due to the presence of clumpy 
regions as seen in the residual images of the cluster (Figure~\ref{fig2}) and cavity 
inflation. However, in the North and South sectors, the metallicity values decreases 
from the centre to the radial distance of \s70 arcsec beyond which the measurements 
are inconclusive due to the less number of data points and poor signal. This suggests
the presence of a shock at the radial distance of \s70 arcsec from the centre of the 
cluster. Similarly, a sharp decrease in pressure is apparent at the same 
locations along the North and South directions, while marginal changes in 
electron number density are observed.

The overall scenario from the profiles shown in Figure~\ref{fig6} manifests that 
the bipolar outflows in the form of jets emerging from the AGN in the central 
dominant galaxy 3C~444 inflates large bubbles (radio lobes or cavities) in the 
ICM of the cluster. These lobes exert pressure on the surrounding gas and 
drive a  shock (vertical dashed lines in Figure~\ref{fig6}) in the ICM 
\citep{2005Natur.433...45M}. The shock features are also seen in the residual 
images in the form of ripples e.g. dark depression surrounding the clumpy 
regions (Figure~\ref{fig2}). This suggests that the radio lobes removed 
substantial amount gas and metals from the cluster centre and spread in 
the ICM at a distance of 100--150 kpc (see metallicity profiles Figure~\ref{fig6}). 
These kind of features were also seen in many clusters such as RBS~797, NGC~5813, 
HCG~62 Abell~2052, ZwCl~2701, Perseus clusters etc. \citep{2011ApJ...737...99B,2006MNRAS.366..417F,2012AdAst2012E...6G,
2009ApJ...697L..95B,2016MNRAS.461.1885V}.

\subsection{Shocks in ICM}

\begin{figure*}
\vbox{
\includegraphics[scale=0.4]{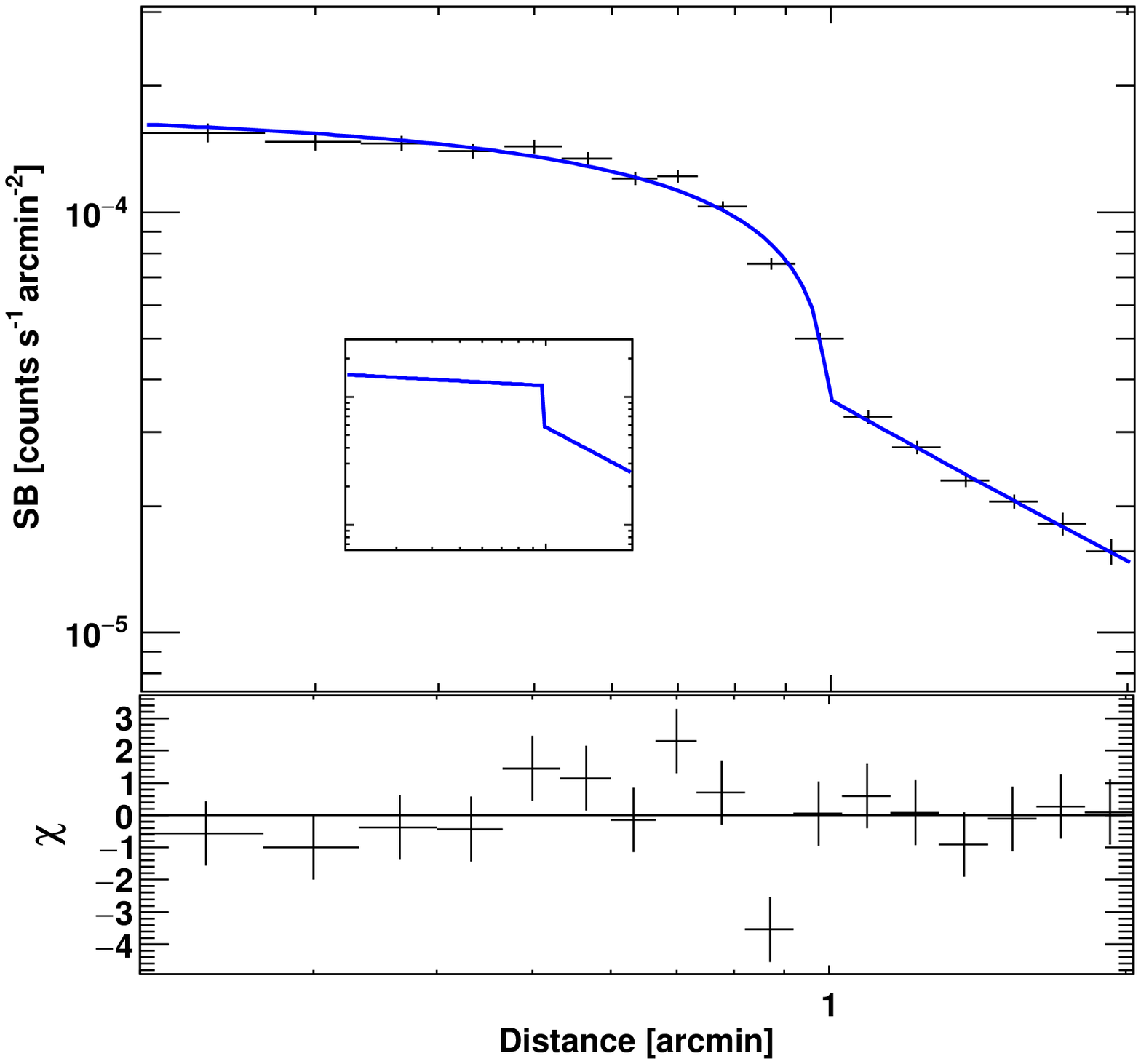}
\includegraphics[trim={2cm 0 0 0},scale=0.4]{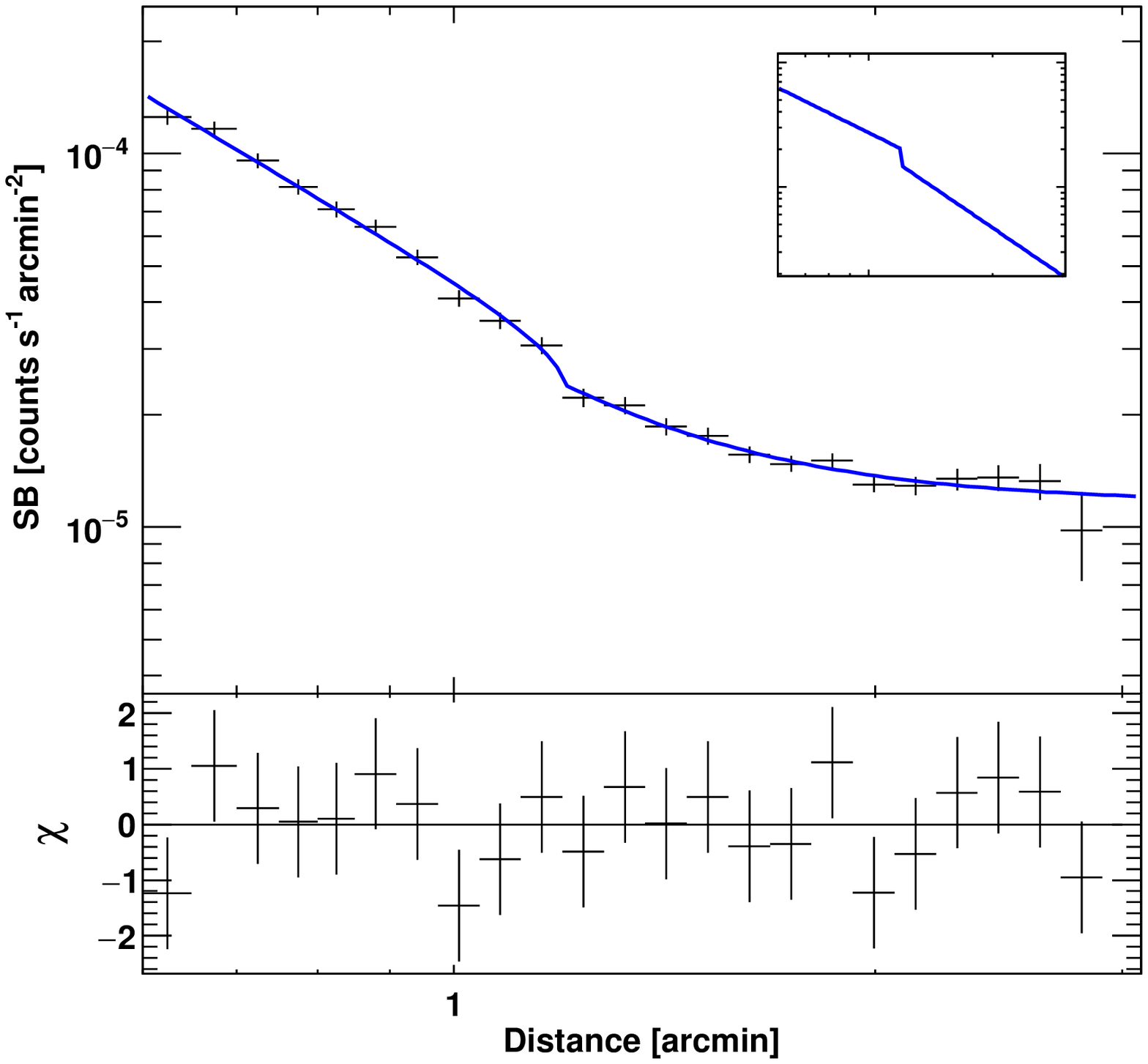}}
\vbox{
\includegraphics[scale=0.4]{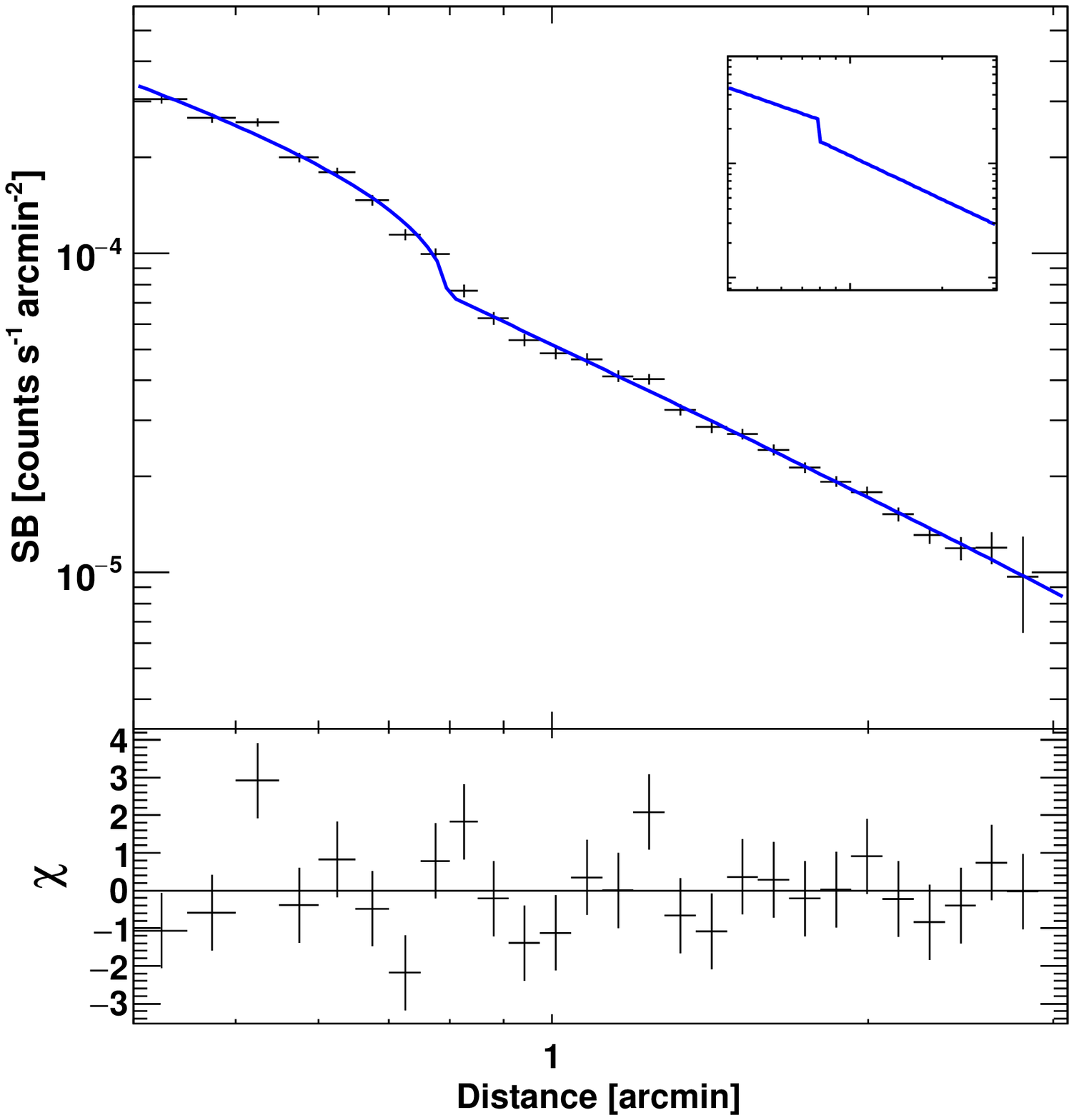}
\includegraphics[trim={2.3cm 0 0 0},clip, width=7cm, height=7.5cm]{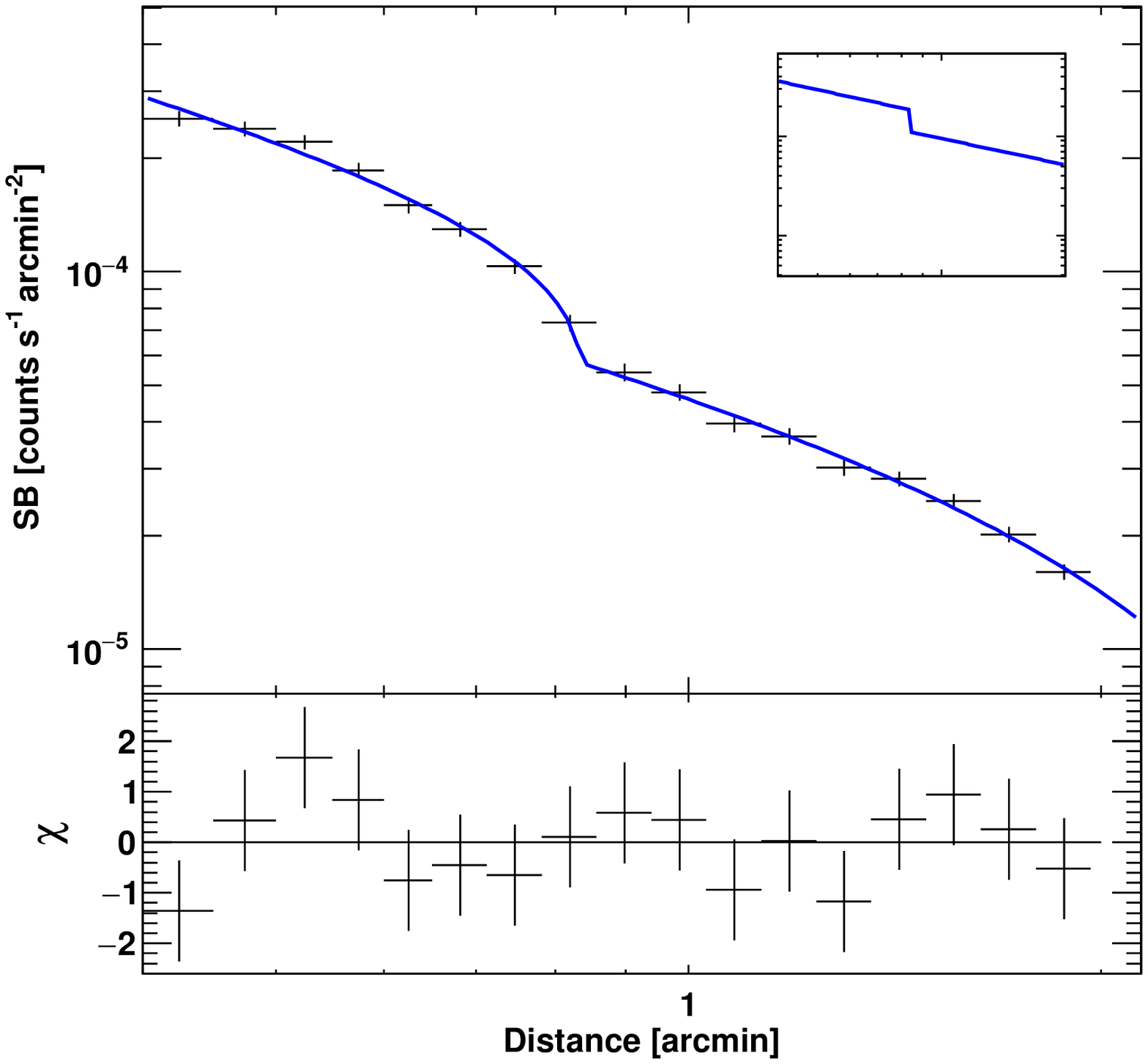}
}
\caption{X-ray surface brightness profiles in 0.3-3 keV band extracted from sectors in
North (25\degr-120\degr -- top left panel), South (215\degr-300\degr -- top right panel), 
East (120\degr-215\degr -- bottom left panel), and West (330\degr-25\degr -- bottom right
panel) directions. The profiles were fitted with broken power-law density model and the 
best fit models are shown in solid lines. Insets in each panel show corresponding 3D 
simulated gas density model. The bottom panels in each figure display the residuals 
obtained from the fitting.}
\label{fig7}
\end{figure*}

The presence of elliptical bright rim-like features in the residual images of A3847 
cluster (Figure~\ref{fig2}) hints the presence of elliptical shock surrounding 
the nuclear region. These shock features are also seen in the sectoral temperature 
profiles in all four directions (Figure~\ref{fig6}), where temperature jumps 
are clearly seen. In order to confirm the shock front in these directions, we 
extracted surface brightness profiles along these directions. The same angular
regions were adopted for all the segments, as mentioned in Section 3.5. We used 
the exposure corrected background subtracted and vignetting corrected image in  
0.3-3.0 keV range. The surface brightness profiles were then fitted with 
the deprojected broken power-law density model within PROFFIT (Version 1.4) package 
\citep{2011A&A...526A..79E}. The density model is defined as:

\begin{eqnarray}
	n(r) = \begin{cases} C\,n_{\rm {0}} \left(\frac{r}{r_{\rm sh}}\right)^{-\alpha1}\,, & \mbox{if } r \le r_{\rm sh} \\ n_{\rm {0}} \left(\frac{r}{r_{\rm sh}}\right)^{-\alpha2}\,, & \mbox{if } r > r_{\rm sh} \end{cases} \,,
\end{eqnarray}

where $n$ is the electron number density which is a function of radius, $n_0$ is the density normalization, $C$ is the density compression factor of the shock, $\alpha$1 and $\alpha$2 
are the power-law indices, $r$ is the radial distance from the centre and $r_{sh}$ is the 
radius corresponding to the putative shock front. All the parameters of the model were allowed 
to vary in the fitting. The best fit broken power-law components along with the surface
brightness profiles are shown in Figure~\ref{fig7}. In North, East and West directions, 
these profiles reveal that there are sudden jumps in the surface brightness values in the 
range of 45--60 arcsec radii. Initially, along the South direction (215\degr - 330\degr\,), 
we did not find a clear change in the profile, possibly due to the variable shock strength in the 
annular region. Therefore, we reduced the angular size of the South segment to 215\degr - 
300\degr\, and fitted the surface brightness profile which showed a minor jump in the 
model at \s70\arcsec\, radius. The best fit parameters obtained from the fitting are tabulated 
in Table~\ref{tab1}. By using an adiabatic index $\gamma=5/3$ and the Rankine-Hugoniot jump
conditions we calculated Mach numbers corresponding to these locations and these are given in the
table. The maximum value of Mach number is  along the North direction and it indicates that the propagation of
the shock is faster compared to that in the East and West directions. The minimum value of Mach number is
along the South direction and it could be due to projection effect, which was also reported earlier
\citep{2011ApJ...734L..28C}. Different values of Mach numbers and positions along different
directions indicate that the observed shock is elliptical (in 2 dimensional figure). 
Similar kind of shocks are also seen in other clusters such as MS~0735 \citep{2005Natur.433...45M,2014MNRAS.442.3192V}, Hydra~A \citep{2009A&A...495..721S,2011ApJ...732...13G}, 
3C~310 \citep{2012ApJ...749...19K}, Virgo \citep{2005ApJ...635..894F,2007ApJ...665.1057F,2010MNRAS.407.2046M}. 

\begin{table*}
\caption{Parameters obtained from the broken power-law density model}
\centering
\small
\begin{tabular}{lllllllllll}\hline
Regions &$\alpha$1& $\alpha$2 & $r_{sh}$ & $n_0$ & C &$\chi^{2}$/dof & Mach No. (${\cal M}$)   \\
&&&(arcmin)&($10^{-4}$)&&     \\ \hline
North(25\degr-120\degr)& $0.13\pm0.04$ &$1.45\pm0.46$ &$0.98\pm0.01$   &$1.20\pm0.05$ &$2.00\pm0.18$  &24.04/11&$1.72\pm0.15$  \\ 
South(215\degr-300\degr)& $1.63\pm0.10$  &$2.25\pm0.48$ &$1.20\pm0.04$ &$0.27\pm0.04$ &$1.36\pm0.15$&12.92/17&$1.25\pm0.10$  \\ 
East(120\degr-215\degr)& $0.94\pm0.08$  &$1.27\pm0.07$ &$0.79\pm0.01$  &$1.94\pm0.15$ &$1.55\pm0.06 $&31.64/21&$1.38\pm0.04$  \\  
West(330\degr-25\degr)& $0.90\pm0.09$   &$0.88\pm0.21$ &$0.83\pm0.01$  &$1.58\pm0.12$ &$1.67\pm0.10$  &10.95/11&$1.47\pm0.07$  \\  \hline
\end{tabular}
\label{tab1}
\end{table*}

\subsection{Different regions in the ICM of 3C~444}

\begin{figure*}
\centering
\includegraphics[width=11cm,height=10cm]{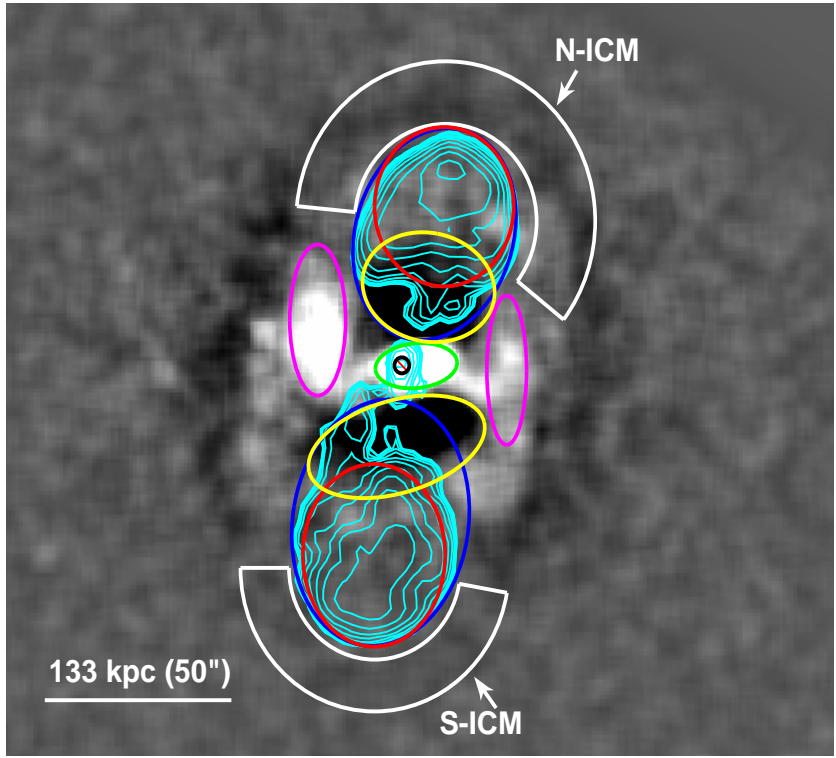} 
\caption{The residual 0.3-6 keV {\it Chandra} image overlaid 
with the 4.89 GHz \textit{VLA} radio contours of 3C~444. The regions, marked 
as ellipsoids in different colours, are selected for the spectral investigation 
of the ICM.}
\label{fig8} 
\end{figure*}

In the earlier section, we described the observed radial thermodynamical properties
and shocked regions in the ICM of A3847. To have a clear understanding of the 
distribution of gas in the central region of the cluster, spectral investigation 
was carried out for various selected regions as marked in Figure~\ref{fig8}. These 
regions include the central compact source (within 2\arcsec\, radius -- marked as 
black circle), central clumpy region (marked as green ellipse, excluding the 
central compact source of 2\arcsec\, radius), clumpy regions in East and West 
directions (marked as magenta ellipsoids), North and South X-ray deficient regions 
(marked as yellow ellipses) and North and South radio bubbles (marked as red 
ellipses). Spectral extraction from these individual regions and model fitting 
were performed as described earlier. Except the central compact source and the central 
clumpy regions, the other regions were well fitted with a single temperature APEC 
and Galactic absorption model. Spectra obtained from nuclear regions required an additional 
power-law component in the model fitting. The cause of the additional component could be due to the 
contribution from the AGN.

Nuclear clumpy region spectrum was initially fitted with a single temperature 
model in which $kT$ and \Zsun\, were allowed to vary. The best-fit parameters obtained
from the single temperature fitting model are given in Table~\ref{tab2}. Reduced $\chi^2$ 
($\chi^2$/dof =148.31/123) of 1.2 suggested that the single temperature model yielded 
a relatively poor fit. We tried to fit the spectrum by using a two components model 
(e.g. APEC+APEC), where the first temperature $kT1$ and abundance $\Zsun 1$ were 
allowed to vary and the second temperature $kT2$ was tied at half of $kT1$ ($kT2=0.5*kT1$) 
and abundance of the second APEC component was tied with the first one e.g. \Zsun 2=\Zsun 1. 
The best fit parameters obtained from this model fitting are given in Table~\ref{tab2}.
Improved value of reduced $\chi^2$ ($\chi^2$/dof=106.62/123) compared to that obtained 
from the single temperature component fitting indicates the presence of cool component in the
core of the cluster.
 
\begin{figure}
\centering
\includegraphics[width=8cm,height=8cm]{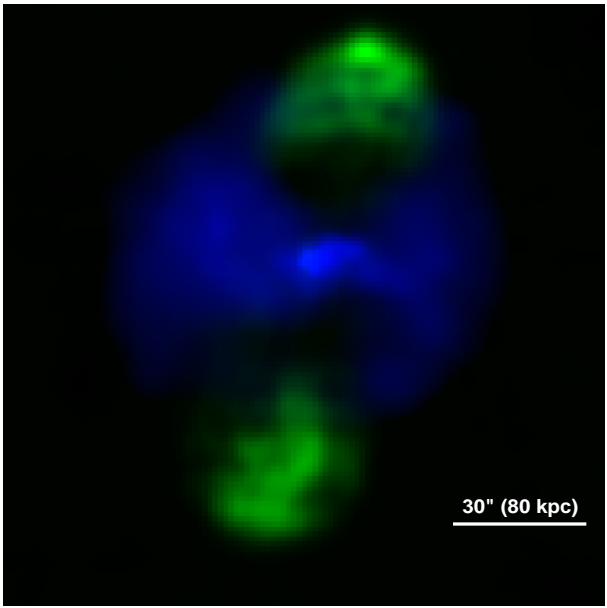} 
\caption{A composite false colour \textit{Chandra} X-ray and \textit{VLA} radio 
emission. Blue and green colours represents the distribution of hot diffuse 
gas (0.3-3 keV range) and 4.89 GHz radio emission map, respectively.}
\label{fig9} 
\end{figure} 

\subsection{X-ray spectra of lobes}

Following detection of non-thermal X-ray emission from the radio lobes in several
FR~II radio sources (\citealt{2005ApJ...626..733C} and references therein), 3C~326 
\citep{2009ApJ...706..454I}, Cyg~A \citep{2010ApJ...714...37Y}, we investigated the presence 
of such component in 3C~444. ICM spectra were extracted by selecting annular regions in the 
surroundings of north and south radio lobes and marked as N-ICM and S-ICM in Figure~\ref{fig8}. 
These spectra were fitted with a single temperature APEC model in which the temperature and 
abundance parameters were allowed to vary. During spectral fitting of S-ICM, as the metal abundance 
reached its lower limit, we froze the parameter to 0.1\Zsun. The best-fit parameters of 
N-ICM and S-ICM spectra are presented in Table~\ref{tab2}.

To investigate the presence of non-thermal X-ray emission associated with the 
radio lobes, we extracted spectra from North-lobe and South-lobe (marked with red ellipses
in Figure~\ref{fig8}). The origin of this non-thermal component is understood to be due 
to the inverse Compton process \citep{2010ApJ...714...37Y}. In the beginning, the spectra
from both the lobes were fitted with a single temperature APEC model. The parameters 
(temperature $kT1$, abundance $Z$, reduced $\chi^2$) obtained from spectral fitting are 
given in Table~\ref{tab2}. The values of North-lobe and South-lobe temperature were found 
to be marginally higher than the surrounding ICM temperature. Though the fitting was 
statistically acceptable, possible presence of power-law component in the lobe emission 
was investigated. Along with the APEC component (with temperature and abundance set at the 
values obtained for surrounding N-ICM and S-ICM), power-law component was added to the fitting model.
Parameters obtained from the two component model fitting to the North and South lobe spectra
are also given in Table~\ref{tab2}. However, addition of non-thermal power-law component to 
the model did not yield significant improvement in the spectral fitting (Table~\ref{tab2}).
Therefore, the non-thermal emission from the radio lobes in 3C~320 is either absent or 
negligible.

\subsection{X-ray and radio morphology}

It is well understood that X-ray deficient regions (cavities) are created by 
powerful radio jets. To investigate the association of the cavities and 
the radio lobes, we generated a false colour image of 4.89 GHz \textit{VLA} radio 
map (Project code-AV0088) and 0.3--3 keV range \textit{Chandra} diffuse image of 
the cluster and is shown in Figure~\ref{fig9}. In this figure, blue and green colours 
represent the hot diffuse gas and 4.89 GHz radio emission, respectively. Though the 
X-ray cavities are formed along the direction of radio jet, physical offset between 
the cavities and radio bubbles (\s61 kpc and \s77 kpc along North and South directions, 
respectively) can be clearly seen in this figure as well as in Figure~\ref{fig8}. 
The observed offset between the X-ray cavities and radio lobes may be either 
due to the projection effect of radio emission or the physical offset between the
X-ray cavities and radio lobes due to two different episodic AGN outburst.
This peculiar nature has been rarely seen in very few other clusters such as
3C~310 \citep{2012ApJ...749...19K}, NGC~1316 \citep{2010ApJ...721.1702L}, 
Abell~4059 \citep{2002ApJ...569L..79H}, MS0735.6+742 \citep{2005Natur.433...45M}. 
The radio jets emitted along North and South directions of 3C~444 were seen 
to bend towards East and West directions at a distance of \s8 arcsec 
(\s20 kpc) from the nuclear region of the central galaxy (Figure~\ref{fig8}). 
Apart from these features, the clumpy hot gas which was located in the core of the
cluster appeared to be shifted towards West direction from the X-ray centre of the 
galaxy. These feature indicated that the system is undergoing some sloshing of hot 
gas because of merging phenomena.

 \section{Discussion}
 
\subsection{Cavity energetics}

It is well known that AGN feedback plays an important role in balancing 
cooling and heating of the ICM \citep{1994ARA&A..32..277F,2005Natur.433...45M,
2004ApJ...607..800B,2006ApJ...652..216R,2012ApJ...753...47D,2012MNRAS.421..808P,
2015Ap&SS.359...61S,2016MNRAS.461.1885V}. Power required to heat the ICM through 
AGN feedback can be derived by taking the ratio between the total energy required to 
create cavities in the ICM and the age of the cavity. Addition of total internal 
energy stored within the cavities and work done by the cavities on the surrounding 
gas of ICM provides the total energy required to create the cavities 
\citep{2004ApJ...607..800B,2006ApJ...652..216R}. Therefore, the total energy required
to create the cavities can be defined as; 
\begin{equation}
E_{cav} = \frac{1}{\gamma_1 - 1}pV + pV = \frac{\gamma_1}{\gamma_1 - 1}pV
\end{equation}
where, $V$ (=4$\pi R_w^2 R_l$/3) is the volume of the ellipsoidal cavity which was 
estimated by visual inspection method \citep{2004ApJ...607..800B,2010ApJ...714..758G,
2016MNRAS.461.1885V}, p (=1.92$n_e$kT) is the pressure of the gas 
surrounding the X-ray cavities and $\gamma$ is the specific heat ratio 
and is equal to 4/3 for relativistic plasma. The parameters required to 
derive the energy of cavities such as volume, surrounding gas pressure, 
temperature and density of gas are given in Table~\ref{tab3}. Thus, the total 
energy ($E_{cav}$) required to create the pair of X-ray cavities in the 
environment of A3847 is estimated to be 3.06$_{-0.79}^{+0.75}$ $\times$ 
10$^{60}$ erg. The age of the cavity was derived by using three different methods 
such as (i) sound crossing time (t$_{sonic}$), (ii) buoyancy time (t$_{buoy}$) and 
(iii) refill time (t$_{refill}$) as discussed in \cite{2004ApJ...607..800B}.
The estimated age by using above three methods are also given in Table~\ref{tab3} 
and are in the range of 5--41 $\times$ 10$^7$ yr.
  
Once we have the cavity energy (E$_{cav}$) and age (t$_{age}$), the total cavity 
power was estimated by using the relation P$_{cav}$ = E$_{cav}$/t$_{age}$. For 
convenience, we assumed that the age estimation through buoyancy equation is a better 
method and the sound crossing time and refill time methods give the lower and upper 
limits (see Table~\ref{tab3}). Hence, the total cavity power was calculated to be 
P$_{cav}$ = 6.13$^{+18.84}_{-2.34}\, \times$ $10^{44}\, erg\, s^{-1}$.

We have also tried to estimate cavity power by assuming that the radio lobes filled 
the X-ray cavities entirely. Though it is not clearly visible in the present case, 
possibly due to the projection effect, the cavity regions were selected by covering 
radio lobes and X-ray deficient regions along north and south directions (blue ellipsoids 
along north and south directions in Figure~\ref{fig8}). Physical parameters for these
two regions were estimated as was done in previous case. These physical parameters 
used for the estimation of cavity power are given as N-cavity (ext) and S-cavity (ext) 
in Table~\ref{tab3} and the total power of the X-ray cavities was found to be 
2.74$_{-1.11}^{+8.54}$ $\times$10$^{45}$ \lum which is approximately 5 times 
higher than the earlier estimate.
  
\subsection{Heating Vs cooling of the ICM}

\begin{figure}
\centering
\includegraphics[width=8cm,height=8cm]{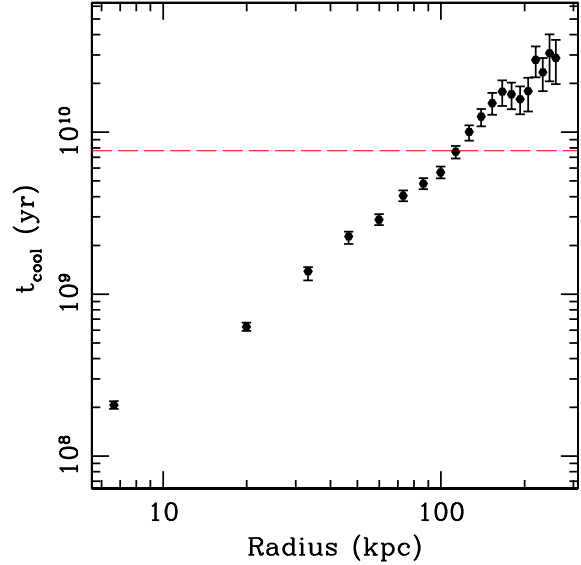} 
\caption{Cooling time profile of the ICM of 3C~444 cluster. The dotted line indicates
the Hubble time.}
\label{fig10} 
\end{figure}

It is believed that in the absence of heating, huge amount of gas in the
ICM cool radiatively and get deposited at the core of the cluster. Time 
required to cool the ICM by radiating its enthalpy was obtained by using 
the expression given in \cite{2009ApJS..182...12C}. The resultant radial 
dependent cooling time profile of the cluster is shown in Figure~\ref{fig10}, 
where the horizontal line at 7.7 Gyr represents the cosmological 
time (or age of the cluster) and corresponding radius is called as the 
cooling radius \citep{2012Natur.488..349M,2004ApJ...607..800B}. The 
cooling radius for A3847 cluster was found to be $\approx$ 100 kpc.
We estimated the cooling power within the cooling radius as 
L$_{cool}$ = 8.30$^{+0.02}_{-0.20}\, \times 10^{43}$ \lum\, and the AGN 
power P$_{cav}$ \s 6.30$\times 10^{44}$ \lum. Comparison of these 
estimates suggested that AGN feedback through radio jets is 
sufficiently large to offset the radiative cooling within the cooling radius. 
 
In the cooling profile (Figure~\ref{fig10}), the cooling time at the 
core of the cluster (within 5\arcsec or 13 kpc) was found to be \s200 
Myr which is \s1/40 of the Hubble time. Even shorter cooling times 
have been reported in several other clusters (e.g. \citealt{2004MNRAS.347.1130V}).
 We then calculated the cooling rate of the gas in ICM using the 
 relation $\frac{2L\mu m_p}{5kT}$ \citep{2012Natur.488..349M,
2004ApJ...607..800B} which was found to be 59.36$^{+1.03}_{-0.43}\,$\Msun\,\pyr\, 
within the cooling radius. This suggested that the radiative energy is being 
replenished on shorter time scale than the cooing time within 13 kpc of 
the core \citep{2005ApJ...625L...9N,2007ARA&A..45..117M}.

{\begin{table*}
\begin{center}
\caption{\label{tab2} Physical parameters of different regions of interest in 3C~444}
\begin{tabular}{@{}lccccr@{}}
\hline 
Regions   & kT1 	    &Photon Index $\Gamma$      		   &Z 		     	&L$_{0.3-10 keV}$   			&$\chi^2/dof$ \\ 
          & (in keV)        &			   &(in Z$_{\odot}$)    &(in 10$^{42}$ erg s$^{-1}$)            &\\
\hline \\
North Cavity  &3.50$_{-0.17}^{+0.21}$    &		   &0.42$_{-0.08}^{+0.09}$	&8.7$_{-0.3}^{+0.2}$	  	&69.5/80 \\ \\
South Cavity  &2.90$_{-0.14}^{+0.14}$    &		   &0.36$_{-0.06}^{+0.07}$	&9.4$_{-0.3}^{+0.2}$		&168.4/151\\ \\
Nucleus 	&1.05$_{-0.03}^{+0.02}$    &1.7$_{-0.1}^{+0.2}$ &0.36$_{-0.08}^{+0.11}$	&1.5$_{-0.1}^{+0.1}$  		&49.7/49 \\ \\
Nuclear Clump 	&1.31$_{-0.02}^{+0.02}$    &		   &0.25$_{-0.03}^{+0.03}$ 	&7.2$_{-0.5}^{+0.5}$      	&148.3/123 \\ \\   
Nuclear Clump$^a$ &2.33$_{-0.07}^{+0.07}$  &		   &0.54$_{-0.05}^{+0.06}$ 	& -- 				&106.6/122 \\
East Clump &3.11$_{-0.13}^{+0.13}$         &		   &0.33$_{-0.06}^{+0.06}$ 	&11.0$_{-0.2}^{+0.2}$ 		&160.4/182 \\ \\
West Clump    &3.62$_{-0.25}^{+0.25}$ &	 		   &0.48$_{-0.11}^{+0.13}$ 	&5.7$_{-0.2}^{+0.2}$ 		&124.6/129 \\ \\
North ICM  &4.41$_{-0.36}^{+0.44}$ &			   &0.23$_{-0.11}^{+0.13}$ 	&6.8$_{-0.4}^{+0.2}$ 		&112.1/104 \\ \\
South ICM  &4.45$_{-0.84}^{+1.72}$ &			   &0.1$^b$			&1.3$_{-0.2}^{+0.1}$ 		&26.7/32 \\ \\
North Lobe &4.43$_{-0.27}^{+0.33}$ &			   &0.21$_{-0.10}^{+0.11}$      &10.0$_{-0.2}^{+0.3}$		&67.9/68\\ \\
South Lobe &5.28$_{-0.38}^{+0.55}$ &			   &0.41$_{-0.15}^{+0.17}$      &8.6$_{-0.2}^{+0.3}$		&62/72\\ \\
North-Lobe$^c$ &4.41$^b$	   &1.9$_{-0.2}^{+0.3}$ &0.23$^b$			&---			        &67.6/68\\ \\
South-Lobe$^c$ &4.45$^b$	   &1.5$_{-0.1}^{+0.1}$ &0.1$^b$			&---			        &64.8/72\\
\hline
\end{tabular}\\
\end{center}
\footnotesize{$^a$ : Data were fitted with APEC+APEC model with temperature of the second APEC component fixed at the half of temperature of first APEC component whereas abundance of second component was tied with that of first component.\\
$^b$ : Corresponding parameters were frozen at the quoted values.\\ 
$^c$ : Presence of power-law (non-thermal) component was tested by adding with the thermal APEC component in the spectral fitting. During 
fitting, the values of temperature and abundance were fixed at corresponding values obtained from the spectral fitting of surrounding ICM.} 
\end{table*}

\begin{table*}
{\renewcommand{\arraystretch}{1.3} 
\centering
\caption{\label{tab3} X-ray cavity energetic parameters of A3847}
\begin{tabular}{@{}lcccr@{}}
\hline
Parameters &  N-cavity & S-cavity & N-cav (ext) & S-cav (ext) \\
\hline
$R_l \times R_w$ (kpc) & 47.5 $\times$ 36.30 & 64.64 $\times$ 31.0 & 70.01 $\times$ 55.63 & 82.52 $\times$ 62.82 \\
$R$ (kpc) & 56.0 & 54.3 & 88.64 & 105.15 \\
Vol ($10^{69}\,$ cm$^{3}$) & 7.67 & 7.62 & 26.59 & 39.97 \\ 

kT (keV) & 4.32$^{+0.37}_{-0.26}$ & 4.19$^{+0.34}_{-0.28}$ & 4.41$^{+0.44}_{-0.36}$ & 4.45$^{+1.72}_{-0.84}$ \\ 
$n_e$ ($10^{-3} \,$ cm$^{-3}$) & 4.93$^{+0.087}_{-0.086}$ & 3.72$^{+0.080}_{-0.078}$ & 8.3$^{+0.19}_{-0.19}$ & 5.4$^{+0.18}_{-0.19}$ \\
$p$ ($10^{-11}$\, erg cm$^{-3}$) & 5.23$^{+1.20}_{-1.20}$ & 4.79$^{+1.11}_{-1.11}$ & 11.24$^{+3.19}_{-3.19}$ & 7.38$^{+4.08}_{-4.08}$ \\
E$_{cav}=4pV$ ($10^{60}\,$ erg) & 1.60$^{+0.37}_{-0.45}$ & 1.46$^{+0.38}_{-0.34}$ & 11.96$^{+3.84}_{-3.39}$ & 11.80$^{+9.73}_{-6.59}$ \\
$C_{sound}$ (km\,s$^{-1}$) & 1054$\pm$31 & 1039$\pm$34 & 1065$\pm$43 & 1070$\pm$101 \\
$v_{cavity}$ (km\, s$^{-1}$) & 352$\pm$73 & 330$\pm$38 & 346$\pm$71  & 337$\pm$70 \\
$t_{sonic}$ ($10^7\,$ yr) & 5.20$\pm$0.16 & 5.12$\pm$0.17 & 8.15$\pm$0.33 & 9.62$\pm$0.90\\
$t_{buoy}$ ($10^7\,$ yr) & 15.56$\pm$3.23 & 16.12$\pm$3.34 & 25.09$\pm$5.2 & 30.51$\pm$6.32 \\
$t_{refill}$ ($10^7\,$ yr) & 40.68$\pm$16.87 & 41.65$\pm$17.27 & 62.83$\pm$26.05 & 73.50$\pm$30.47 \\
\hline
\end{tabular}}
\footnotesize
\end{table*}


\subsection{Central engine and the super massive black hole}

If the AGN feedback is the primary source of ICM gas heating, then it is 
important to study the properties of the central engine of the galaxy 
cluster. It is well understood that every central dominated bright 
galaxy in clusters harbour a super massive black hole (SMBH) of mass
in the range of 10$^6$ -- 10$^{9}$ \Msun. Such SMBHs accrete matter 
from the surrounding and releases huge amount of energy in the form of
radiation and/or outflows \citep{1970Natur.226...64B,2009Natur.460..213C}. 
Even a small fraction (less than 1$\%$) of the released energy is sufficient 
to heat the bulge \citep{2009Natur.460..213C} of the galaxy. Therefore, 
to study the properties of the central engine (AGN), it is necessary to 
understand the accretion and emission processes in the SMBH. For this, 
we derived mass of the black hole located at the centre of the bright 
galaxy in A3847 using stellar dispersion velocity and K-band 
luminosity. The stellar dispersion velocity $\sigma$ = 164\pms\,34 km 
s$^{-1}$ (as quoted by \citealt{1990ApJ...356..399S}) was used in our 
calculation, whereas the K-band luminosity (L$_K$ = 1.47\pms 0.06 
$\times 10^{44}$\lum) was taken from the NASA Extragalactic Database 
(NED). The mass of the black hole was estimated by using dispersion 
velocity and K-band luminosity methods to be 6.07\pms1.8 \tim$10^7$\Msun\, 
and 7.09 \pms0.60 \tim$10^7$ \Msun, respectively, and found to agree 
within one sigma limits. Mass accretion rate ($\dot{M}_{acc}$) of the SMBH 
in 3C~444 was obtained by using the relation $\dot{M}_{acc}$ = $P_{cav}/\epsilon \times c^2$, 
where $P_{cav}$ is cavity power and $c$ is the velocity of light. 
In the present work, the value of radiative efficiency $\epsilon$ 
is assumed to be 0.1 \citep{2006ApJ...652..216R}. Using this 
relation, the accretion rate was estimated to be $\dot{M}_{acc}$ = 
0.11$_{-0.04}^{+0.14}$ \Msun\,\pyr.

Eddington accretion rate of a black hole depends on its mass and  
radiative efficiency through the relation $\dot{M}_{Edd}$ = 
$L_{Edd}/\epsilon \times c^2$, where $L_{Edd}$ is the Eddington 
luminosity. For the SMBH in the central dominant galaxy 3C~444, 
the Eddington accretion rate was estimated to be $\dot{M}_{Edd}$ = 
1.34\pms 0.26 \Msun\,\pyr. We estimated Bondi accretion rate by 
assuming spherically symmetric and steady state accretion onto the 
compact object. Bondi accretion process occurs within the Bondi 
radius and is defined as R$_{B}$ = 2G$M_{BH}/c_s^2$. The Bondi 
accretion rate for compact object is given by $\dot{M}_{B} = \pi 
R_B^2 \rho c_s$, where  $c_s$ is sound speed in the surrounding 
medium. It is difficult to estimate the gas density within Bondi 
radius due to the limitations in the angular resolution of detectors.
However, we used the gas density obtained from extraction of 
nuclear 2\arcsec spectrum, though it underestimate the value of
number density. Using the mass of the SMBH in the bright galaxy in
A3847 cluster, Bondi radius was estimated to be 1.79$_{-0.30}^{+0.31}$ \tim 
10$^{-3}$ kpc. By using the value of number density obtained from 
nuclear 2\arcsec spectrum (n$_e$ = 0.075$\pm$0.007 $cm^{-3}$) and 
estimated Bondi radius, the Bondi accretion rate was derived to be 
$\dot{M}_{B}$ = 3.08$_{-1.41}^{+1.43}$ \tim 10$^{-6}$ \Msun\,\pyr.

The ratios of mass accretion onto the SMBH in 3C~444 ($\dot{M}_{acc}$) 
to the Bondi and Eddington accretion rates were found to be 
3.5$_{-3.0}^{+6.4}$\tim\,10$^4$ and 0.08$_{-0.05}^{+0.12}$, 
respectively. The accretion rate is supposed to be sub-Eddington 
when $\dot{M}_{acc}/\dot{M}_{Edd}$ is in the range of 10$^{-4} - 10^{-2}$,
where AGNs are often observed to be radiatively inefficient 
\citep{2013MNRAS.432..530R}. \cite{2015A&A...579A..62G} suggested 
that black hole accretion rate can be boosted up to 10$^{-2}$ times the
Eddington rate and two to three orders above the Bondi accretion rate via 
a process known as ``chaotic cold accretion'' (CCA). In CCA model the cold 
 filament and clouds condense out of the hot gas phase where cooling time 
 is much low (in Myr order within 10 kpc region) and rain onto the SMBH, 
 boosting the feedback mechanism and re-heating the ICM core.

\section{Conclusions}

Using the deep {\it Chandra} observation and 4.89 GHz VLA radio map,
we investigated the properties of X-ray deficient regions, shocks
and other substructures in the ICM of A3847 cluster. The results obtained
from our study are summarized as follows.

\begin{enumerate}
\item[1] A pair of giant X-ray deficient cavities along North and South directions
were clearly detected in the residual images of the cluster.

\item[2]X-ray and 4.8 GHz radio images revealed the peculiar positioning 
of the cavities and radio bubbles in A3847. The radio lobes and X-ray
cavities are apparently not spatially coincident and they exhibit offset, by 
\s61 kpc and \s77 kpc from each other along the North and South directions, respectively.

\item[3] Presence of shock is detected in the ICM of cluster A3847. 

\item[4]Using results obtained from the radial thermodynamical profiles, it 
is confirmed that the radio bubbles are responsible for removing substantial 
amount of matter from the centre of the cluster.

\item[5] The AGN feedback power that heats up the ICM was estimated to be 
\s6.3 $\times 10^{44}$ \lum\, and is sufficiently large to offset the cooling 
(L$_{cool}$ \s\, 8.30 $\times\, 10^{43}$ \lum) of the ICM in A3847 cluster.

\item[6] The ratio of estimated accretion rate to the Eddington rate and 
Bondi rate suggested that the SMBH in the central dominant galaxy 3C~444
accretes matter through chaotic cold accretion.
\end{enumerate}

\section*{Acknowledgments} 
The authors are grateful to the anonymous referee for constructive comments and 
suggestions on the paper. The authors thank A. R. Rao for careful reading of the 
manuscript. This work has made use of data from the \textit{Chandra}, \textit{VLA} 
and Gemini-South archive, NASA's Astrophysics Data System(ADS), Extragalactic Database 
(NED), software provided by the \textit{Chandra} X-ray Center (CXC), HEASOFT for 
spectral fitting and Veusz plotting tools.

\def\aj{AJ}%
\def\actaa{Acta Astron.}%
\def\araa{ARA\&A}%
\def\apj{ApJ}%
\def\apjl{ApJ}%
\def\apjs{ApJS}%
\def\ao{Appl.~Opt.}%
\def\apss{Ap\&SS}%
\def\aap{A\&A}%
\def\aapr{A\&A~Rev.}%
\def\aaps{A\&AS}%
\def\azh{AZh}%
\def\baas{BAAS}%
\def\bac{Bull. astr. Inst. Czechosl.}%
\def\caa{Chinese Astron. Astrophys.}%
\def\cjaa{Chinese J. Astron. Astrophys.}%
\def\icarus{Icarus}%
\def\jcap{J. Cosmology Astropart. Phys.}%
\def\jrasc{JRASC}%
\def\mnras{MNRAS}%
\def\memras{MmRAS}%
\def\na{New A}%
\def\nar{New A Rev.}%
\def\pasa{PASA}%
\def\pra{Phys.~Rev.~A}%
\def\prb{Phys.~Rev.~B}%
\def\prc{Phys.~Rev.~C}%
\def\prd{Phys.~Rev.~D}%
\def\pre{Phys.~Rev.~E}%
\def\prl{Phys.~Rev.~Lett.}%
\def\pasp{PASP}%
\def\pasj{PASJ}%
\def\qjras{QJRAS}%
\def\rmxaa{Rev. Mexicana Astron. Astrofis.}%
\def\skytel{S\&T}%
\def\solphys{Sol.~Phys.}%
\def\sovast{Soviet~Ast.}%
\def\ssr{Space~Sci.~Rev.}%
\def\zap{ZAp}%
\def\nat{Nature}%
\def\iaucirc{IAU~Circ.}%
\def\aplett{Astrophys.~Lett.}%
\def\apspr{Astrophys.~Space~Phys.~Res.}%
\def\bain{Bull.~Astron.~Inst.~Netherlands}%
\def\fcp{Fund.~Cosmic~Phys.}%
\def\gca{Geochim.~Cosmochim.~Acta}%
\def\grl{Geophys.~Res.~Lett.}%
\def\jcp{J.~Chem.~Phys.}%
\def\jgr{J.~Geophys.~Res.}%
\def\jqsrt{J.~Quant.~Spec.~Radiat.~Transf.}%
\def\memsai{Mem.~Soc.~Astron.~Italiana}%
\def\nphysa{Nucl.~Phys.~A}%
\def\physrep{Phys.~Rep.}%
\def\physscr{Phys.~Scr}%
\def\planss{Planet.~Space~Sci.}%
\def\procspie{Proc.~SPIE}%
\let\astap=\aap
\let\apjlett=\apjl
\let\apjsupp=\apjs
\let\applopt=\ao
\bibliographystyle{mn2e}
\bibliography{mybib}




\end{document}